\documentclass[useAMS,usenatbib,onecolumn]{mn2e}
\usepackage{epsfig,amssymb,rotating,lscape,graphicx}
\newcommand{\D}{\displaystyle}

\def\eg{{e.g.\ }}
\def\ar{{ARA\&A} \,}
\def\aa{{A\&A} \,}
\def\apj{{ApJ} \,}
\def\apjs{{ApJS} \,}
\def\n{{Nat} \,}
\def\mn{{MNRAS} \,}

\def\na{{NewA} \,}

\def\nrel{nonrelativistic\,\,}

\title[Impact of a non-Gaussian density field on Sunyaev-Zeldovich observables]
{Impact of a non-Gaussian density field on Sunyaev-Zeldovich observables}

\author[S. Sadeh, Y. Rephaeli, J. Silk]{Sharon Sadeh$^{1}$\thanks{E-mail:
shrs@post.tau.ac.il}, Yoel Rephaeli$^{1,2}$ and Joseph 
Silk$^{3}$\\
$^{1}$School of Physics and Astronomy, Tel Aviv University, Tel Aviv, 69978, Israel\\
$^{2}$Center for Astrophysics and Space Sciences, University of California, 
San Diego, La Jolla, CA 92093-0424\\
$^{3}$Department of Astrophysics, University of Oxford, Keble Road, OX1 3RH, UK}

\begin{document}


\pagerange{\pageref{firstpage}--\pageref{lastpage}} \pubyear{2006}

\maketitle

\label{firstpage}

\begin{abstract}
The main statistical properties of the Sunyaev-Zeldovich (S-Z) effect - 
the power spectrum, cluster number counts, and angular correlation 
function - are calculated and compared within the framework of two 
density fields which differ in their predictions of the cluster mass 
function at high redshifts. We do so for the usual Press \& Schechter 
mass function, which is derived on the basis of a Gaussian density 
fluctuation field, and for a mass function based on a $\chi^2$ 
distributed density field. These three S-Z observables are found to be 
very significantly dependent on the choice of the mass function. The 
different predictions of the Gaussian and non-Gaussian density fields 
are probed in detail by investigating the behaviour of the three S-Z 
observables in terms of cluster mass and redshift. The formation time 
distribution of clusters is also demonstrated to be sensitive to the 
underlying mass function. A semi-quantitative assessment is given of 
its impact on the concentration parameter and the temperature of 
intracluster gas.
\end{abstract}

\begin{keywords}
galaxies:clusters:general -- cosmic microwave background --
large-scale structure of the Universe
\end{keywords}

\section{INTRODUCTION}

It has recently been suggested that several observational results
may be hinting at a significant non-Gaussian component in the 
primordial density fluctuation field (Mathis, Diego, \& Silk 2004, 
hereafter MDS). Among these are the detection of structures with high 
velocity dispersions  at redshifts $z=4.1$ (Miley et al. 2004) and $z=2.1$ 
(Kurk et al. 2004), the apparent slow evolution of the X-ray cluster 
luminosity function indicated by objects taken from the Vikhlinin et al. 
(1998) and Mullis et al. (2003) cluster catalogs, and the enhancement 
of the CMB power spectrum at high multipoles that was measured by the 
\textit{CBI} (Mason et al. 2003) and \textit{ACBAR} (Kuo et al. 2004) 
experiments. MDS pointed 
out that scale-dependent non-Gaussianity may be probed through its 
detectable manifestations on the large mass scales of clusters of 
galaxies, and that this is possible without necessarily violating 
constraints on non-Gaussianity such as those obtained from \textit{WMAP} 
all-sky CMB maps (Komatsu et al. 2003). It is argued by MDS that while 
it is hard (although possible) to reconcile the observational 
results with predictions made on the basis of a Gaussian field, 
these arise more naturally with either positively-or 
negatively-skewed density fields. 

In this paper we extend the work of MDS, focusing on the impact 
of a positively-skewed non-Gaussian field on the principal 
statistical properties of the Sunyaev-Zeldovich (S-Z) effect. We 
compare in detail the predicted S-Z power spectrum, cluster number 
counts, and the angular correlation function for Gaussian and 
non-Gaussian models. 

In section 2 we specify the various models and scalings used in 
order to carry out the calculations; specifically, a complete 
account of the mass functions and the relevant scalings for the 
quantification of cluster-induced Comptonization are provided. 
Results of the calculations relating to the S-Z power spectrum and 
number counts are detailed in section 3. In section 4 we formulate 
the method of calculation of the 2-point angular correlation 
function of S-Z clusters and present the corresponding results. 
The calculation of cluster formation times within the framework 
of non-Gaussian fields requires a slight modification of the 
corresponding Gaussian case. This is explained in section 5. 
Results of the formation time computations in both the Gaussian 
and non-Gaussian models follow in the same section, followed by 
a brief discussion and a summary of our main conclusions in 
section 6.

\section{The Model}

Calculations of statistical properties of the evolving cluster 
population involve the basic cosmological and large-scale quantities 
and parameters, as well as essential properties of intracluster (IC) 
gas. Detailed description of the calculation of integrated S-Z 
properties can be found in the literature. Therefore, our brief 
treatment here highlights only those aspects that pertain to the 
basic properties of the density fields explored in this work.
The multiplicity function in the Press \& Schechter (1974) formalism is 
\begin{equation}
n(M,z)=-F(\mu)\frac{\rho_b}{M\sigma}\frac{d\sigma}{dM}dM,
\label{eq:nmz}
\end{equation}
where $\mu\equiv\delta_c/\sigma$ and $\rho_b$ is the background density. 
The critical overdensity for spherical collapse is assumed to be constant, 
$\delta_c=1.69$, owing to its weak dependence on redshift. The density 
field variance, smoothed over a top-hat window function of radius $R$, 
is calculated through
\begin{equation}
\sigma^{2}(R)=\int_0^{\infty}P(k)\widetilde{W}^2(kR)k^2dk,
\end{equation}
where $P(k)\equiv Ak^{n}T^{2}(k)$. The mass variance evolves with 
redshift according to
\begin{equation}
\sigma(M,z)=\frac{\sigma(M,0)}{1+z}
\frac{g[\Omega_m(z)]}{g[\Omega_m(0)]},
\end{equation}
where
\begin{equation}
\Omega_m(z)=\frac{\Omega_m(0)(1+z)^3}{(1+z)^2[1+\Omega_m(0)z-
\Omega_{\Lambda}(0)]}; \,\,\,\Omega_{\Lambda}(z)=1-\Omega_m(z),
\end{equation}
and (Carroll, Press \& Turner 1992)
\begin{equation}
g[\Omega_m(z)]=\frac{2.5\Omega_m(z)}{[\Omega_m(z)^{4/7}-\Omega_{\Lambda}+
(1+\Omega_m(z)/2)(1+\Omega_{\Lambda}/70)]}. 
\end{equation}
The function $F(\mu)$ in equation~(\ref{eq:nmz}) assumes the form 
\begin{equation}
F(\mu)=\sqrt{\frac{2}{\pi}}\D{e^{-\frac{\mu^2}{2}}}\frac{\mu}{\sigma_R},
\label{eq:f_g}
\end{equation}
for a Gaussian density field, and the form 
\begin{equation}
F(\mu)=\frac{\left(1+\sqrt{\frac{2}{m}}\mu\right)^{m/2-1}}{(\frac{2}{m})
^{(m-1)/2}\Gamma{(\frac{m}{2})}}e^{\left[-\frac{m}{2}\left(1+\sqrt
{\frac{m}{2}}\mu\right)\right]}\frac{\mu}{\sigma_R},
\label{eq:f_ch}
\end{equation}
for a density fluctuation field which is distributed according to the
$\chi^2_m$ model (Koyama, Soda \& Taruya 1999). The parameter $m$ 
describes the number of CDM fields added in quadratures, so as to 
yield the requested distribution. 

In most of our calculations we use $m=1$, which constitutes the 
largest possible deviation from a normally distributed probability 
density function (PDF) for this model. Increasing the number of 
random density fields obviously results in a PDF that approaches a 
Gaussian, in accordance with the central limit theorem. Some of the 
calculations were done also for the case $m=2$ in order to illustrate 
the impact of this parameter on the high-mass tail of the PDF. The 
respective mass functions are then obtained by multiplying either 
expression by a common factor, as explicitly stated in 
equation~(\ref{eq:nmz}). 

For the Gaussian case an adiabatic CDM transfer function is employed,
\begin{equation}
T(k)_{CDM}=\frac{\ln{(1+2.34q)}}{2.34q}[1+3.89q+(16.1q)^2+(5.46q)^3+
(6.71q)^4]^{-1/4},
\end{equation}
whereas an isocurvature CDM transfer function is chosen for the 
$\chi^{2}_m$ model,
\begin{equation}
T(k)_{CDM,isoc}=(5.6q)^{2}\left[1+\frac{(40q)^2}{1+
215q+(16q)^2(1+0.5q)^{-1}}+(5.6q)^{8/5}\right]^{-5/4},
\end{equation}
with $q\equiv k/(\Omega_m h^2) \, Mpc^{-1}$. Both transfer functions were 
taken from Bardeen et al. (1986). The calculations were carried out for a 
$\Lambda$CDM model, with $\Omega_{\Lambda}=0.7$, $\Omega_m=0.3$, $h=0.7$. 
In the Gaussian case the spectral index was taken to be $n=1$ and the 
normalization $\sigma_8=0.9$. In the non-Gaussian case $n=-1.8$, and from 
the requirement that the present cluster abundance is the same as 
calculated in the Gaussian model, we obtain $\sigma_8=0.73$.

The S-Z angular power spectrum is calculated using the basic expression 
\begin{equation}
C_{\ell}=\int_z\,r^2\frac{dr}{dz}\int_M n(M,z)\,G_{\ell}(M,z)\,dM\,dz,
\label{eq:clmb}
\end{equation}
where $r$ is the co-moving radial distance, and $G_{\ell}$ is obtained 
from the angular Fourier transform of the profile of the temperature change 
induced by Comptonization of the CMB at an angular distance $\theta$ from 
the centre of a cluster, $\Delta T(\theta)$. The function $G_{\ell}$, which 
is proportional to $\Delta T(\theta)^2$, is fully specified in, e.g., 
Molnar \& Birkinshaw (2000). In the limit of non-relativistic electron 
velocities (more on this in section 3.2) the relative temperature change  
assumes the simple form 
\begin{equation}
\frac{\Delta T}{T}(\theta)=\left[x\coth{\left(\frac{x}{2}\right)}-4\right]
y(\theta),
\end{equation}  
where, for an isothermal cluster with a $\beta$ density profile,
the Comptonization parameter is 
\begin{eqnarray}
y(\theta)=2\frac{k_B\sigma_T}{m_e c^2} \frac{n_0(M,z)T_0(M,z)r_c(M,z)} 
{\sqrt{1+(\theta/\theta_c)^2}}\tan^{-1}\left[p\sqrt{\frac{1-
(\theta/p\theta_c)^2}{1+(\theta/\theta_c)^2}}\right].
\label{eq:ythet}
\end{eqnarray}
Here $n_0$, $T_0$, $r_c$, and $\theta_c$ denote the central electron 
density, temperature, core radius, and the angle subtended by the core, 
respectively. The virial radius is taken throughout to be $p=10$ times 
$r_c$.

Of crucial importance are the scalings of cluster properties with mass and 
redshift; the temperature is assumed to scale as
\begin{equation}
T(M,z)=T_{15}\left(\frac{M}{10^{15}h^{-1}M_{\odot}}\right)^{\alpha}
(1+z)^{\psi},
\label{eq:tempsc}
\end{equation}
where a temperature of $8.5\,keV$ is ascribed to a cluster with mass 
$10^{15} h^{-1}M_{\odot}$ at present. The parameter $\alpha$ is usually 
taken to be $2/3$, in accordance with theoretical predictions based on 
hydrostatic equilibrium. We allow a variance of $10\%$ in its value, 
as to possibly address non-gravitational effects which may result in a 
deviation from the simplistic assumption of hydrostatic equilibrium, 
and also to reflect the observed variance in this parameter. An 
additional uncertainty relating to this scaling is represented in the 
parameter $\psi$, which describes the temporal evolution of cluster 
temperatures, and is taken to be either zero or unity, constituting two 
limiting cases for which there is either no evolution, or strong 
evolution, respectively. Little is known on the evolution of the X-ray 
temperature function due to the lack of X-ray observations of 
high-redshift clusters, although some interesting insight is gained 
from hydrodynamical simulations (Norman 2005).

The electron density is parameterized as
\begin{equation}
n_e(M,z)\simeq n_0\frac{f}{0.1}(1+z)^{3}, 
\end{equation}
with the scaling of the gas mass fraction to $\sim 10\%$ (\eg, Carlstrom, 
Holder \& Reese 2002). Since we do not yet know the redshift dependence 
of $f$, we assume $f=0.1$ to be roughly valid throughout the redshift 
considered here. Finally, the core radius is calibrated according to the 
simple relation
\begin{equation}
r_c=0.15h^{-1}\,\left(\frac{M}{10^{15}h^{-1}M_{\odot}}\right)^{1/3}\frac{1}{1+z}
\,Mpc.
\end{equation}
In accord with observational results, we adopt a variance of $20\%$ in the 
value of the core radius. In order for the mass to remain constant when 
different scalings are used for the core radius, a corresponding change 
is affected in the central gas density.

The number of clusters with S-Z flux (change) above 
$\Delta \overline{F}_{\nu}$ is (e.g. Colafrancesco et al. 1997)
\begin{equation}
N(>\Delta \overline{F}_{\nu})=\int r^{2}\frac{dr}{dz}dz 
\int_{\Delta \overline{F}_{\nu}}n(M,z)\,dM.
\label{eq:nc}
\end{equation}
For a cluster with mass M at redshift z 
\begin{equation}\Delta F_{\nu}=\D\frac{2(k_{b}T)^{3}}{(hc)^{2}}g(x)y_{0}
\int R_{s}(|\hat{\gamma}-\hat{\gamma_{\ell}}|,\sigma_{B})\cdot
y(|\hat{\gamma_{\ell}}|,M,z)\,d\Omega,
\end{equation}
where $\hat{\gamma_{\ell}}$ and $\hat{\gamma}$ denote line of sight (los) 
directions through the cluster centre, and relative to this central los, 
respectively, 
\begin{equation}
y_0=2\frac{k_b\sigma_T}{m_e c^2}n_0(M,z)T_0(M,z)r_c(M,z),
\end{equation}
and the spectral dependence (of the thermal component) of the effect is 
given (in the \nrel limit; more on this in section 3.2) by 
\begin{equation}
g(x)\equiv\frac{x^4e^{x}}{(e^{x}-1)^2}\left[x\coth{(x/2)-4}\right] , 
\end{equation}
where $x=h\nu/kT$.
The profile of the effect is given in 
\begin{equation}
y(|\hat{\gamma_{\ell}}|,M,z)\equiv\D\frac{1}{\sqrt{1+(\theta/\theta_{c})^{2}}}
\cdot\tan^{-1}
{\left[p\sqrt{\frac{1-(\theta/p\theta_{c})^{2}}{1+(\theta/\theta_{c})^{2}}}\right]},
\end{equation}
which is the los integral along a direction that forms an angle $\theta$ 
with the cluster centre. $R_s(|\hat{\gamma}-\hat{\gamma_{\ell}}|,
\sigma_{B})$ denotes the angular response of a detector whose beam size 
is given in terms of $\sigma_B$. Finally, 
\begin{equation}
\Delta\overline{F}_{\nu}=\D\frac{\int \Delta F_{\nu}E(\nu)d\nu}{\int E(\nu)d\nu},
\end{equation}
is the flux weighted over the detector spectral response $E(\nu)$. 

Only clusters for which the S-Z flux exceeds the \textit{PLANCK/HFI} detection 
limit of $30\,mJy$ are considered here. This is manifested in the lower 
limit of the mass function integral, equation~(\ref{eq:nc}). The results 
for the S-Z power spectrum and number counts described below refer to 
these two models: (a) IC gas temperature evolves with time according to 
$\psi=1$, and (b) no temperature evolution, $\psi=0$. Power spectra 
were calculated at $\nu =353$ GHz, with a beam size is $7.1'$. Number 
counts were calculated also at $\nu=143$ GHz and $\nu=545$ GHz.

\section{Results}

\subsection{Power spectrum}

Several qualitative assessments can be pointed out before we describe 
specific results. The longer tail of the non-Gaussian distribution 
function (with respect to that of the Gaussian) at the high mass end 
translates to higher cluster abundance in general, and high mass clusters 
in particular. This holds in the entire redshift range (with the 
exception of $z=0$, since the mass function in the non-Gaussian case 
was normalized such as to yield the same cluster abundance as in the 
Gaussian case). This tendency stems from the fact that a 
positively-skewed density fluctuation field exhibits earlier 
cluster formation by enhancing the probability that the average 
density within a spherical region of any size exceeds the critical density 
for collapse. Consequently, the S-Z power spectrum is expected to attain 
higher levels in the non-Gaussian case. Owing to the inferred denser 
population of clusters at high redshifts, the magnitude at the peak should 
shift to higher multipoles, reflecting the higher density of distant 
clusters with smaller angular sizes. Naturally, this effect would be 
drastically reduced in a non-evolving temperature scenario due to the 
inferred lower temperatures associated with high redshift clusters when 
the ($1+z$) scaling is factored out. 

\begin{figure*}
\centering
\epsfig{file=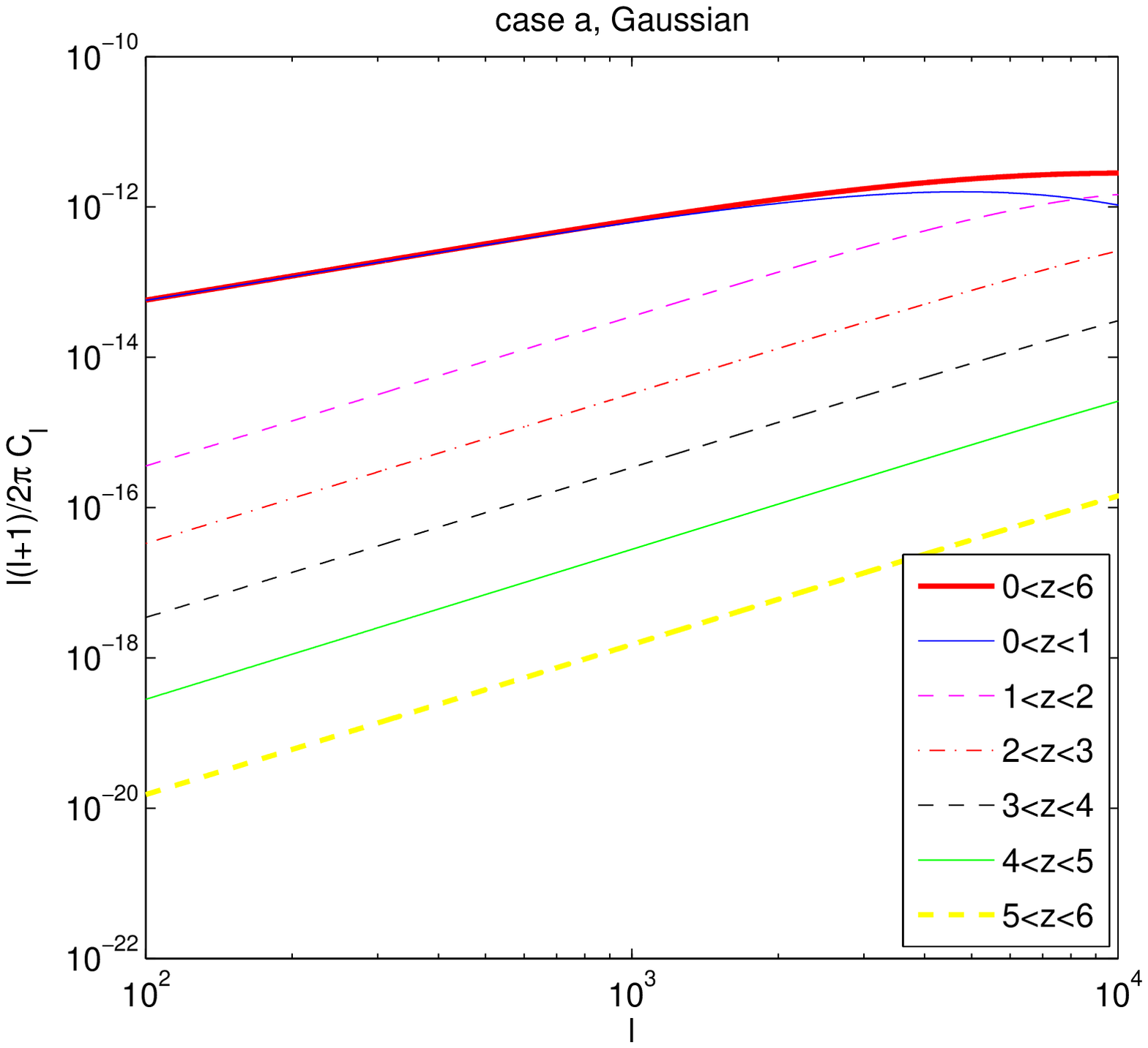, height=7.cm, width=8cm,clip=}
\epsfig{file=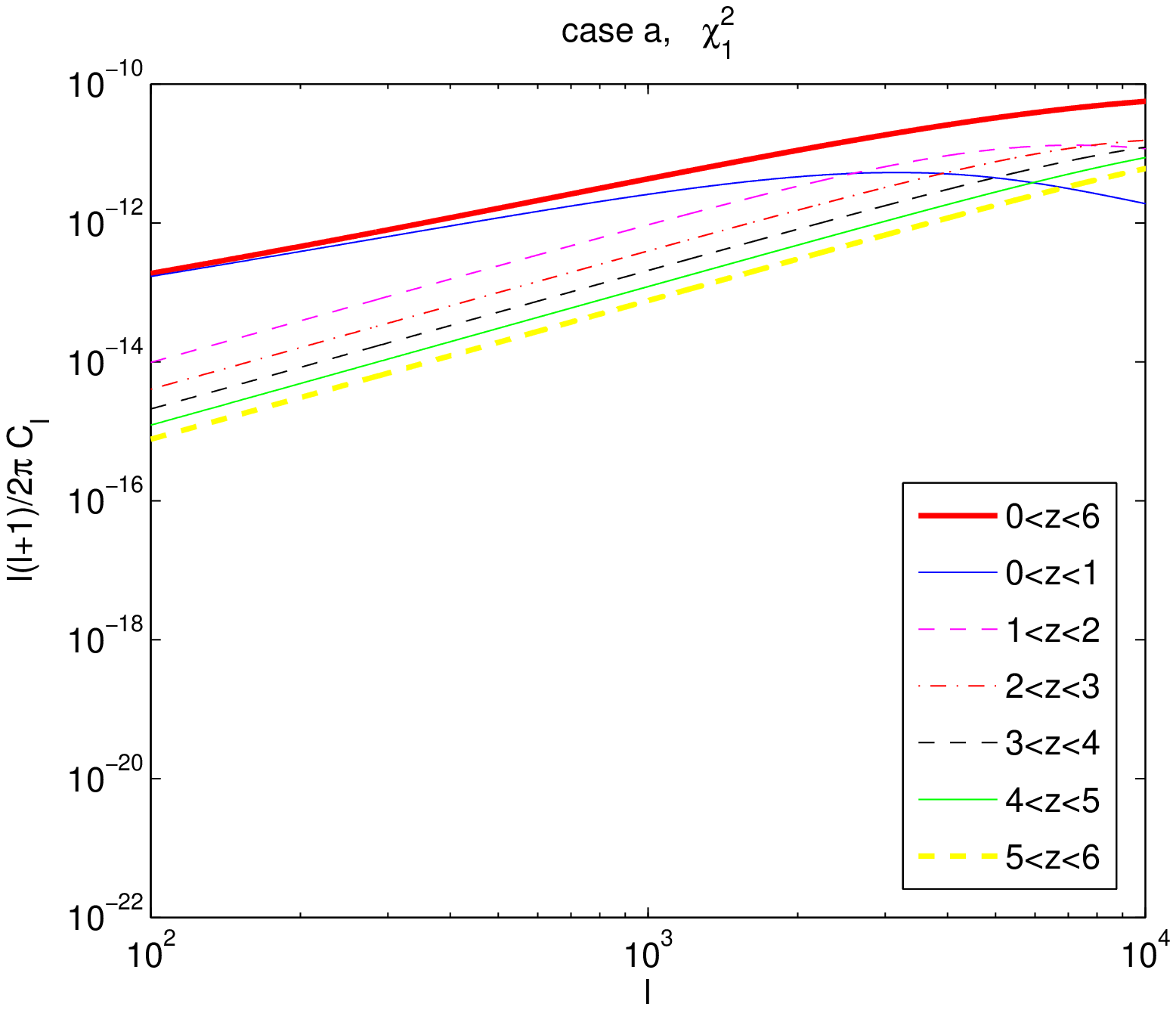, height=7.cm, width=8cm,clip=}
\epsfig{file=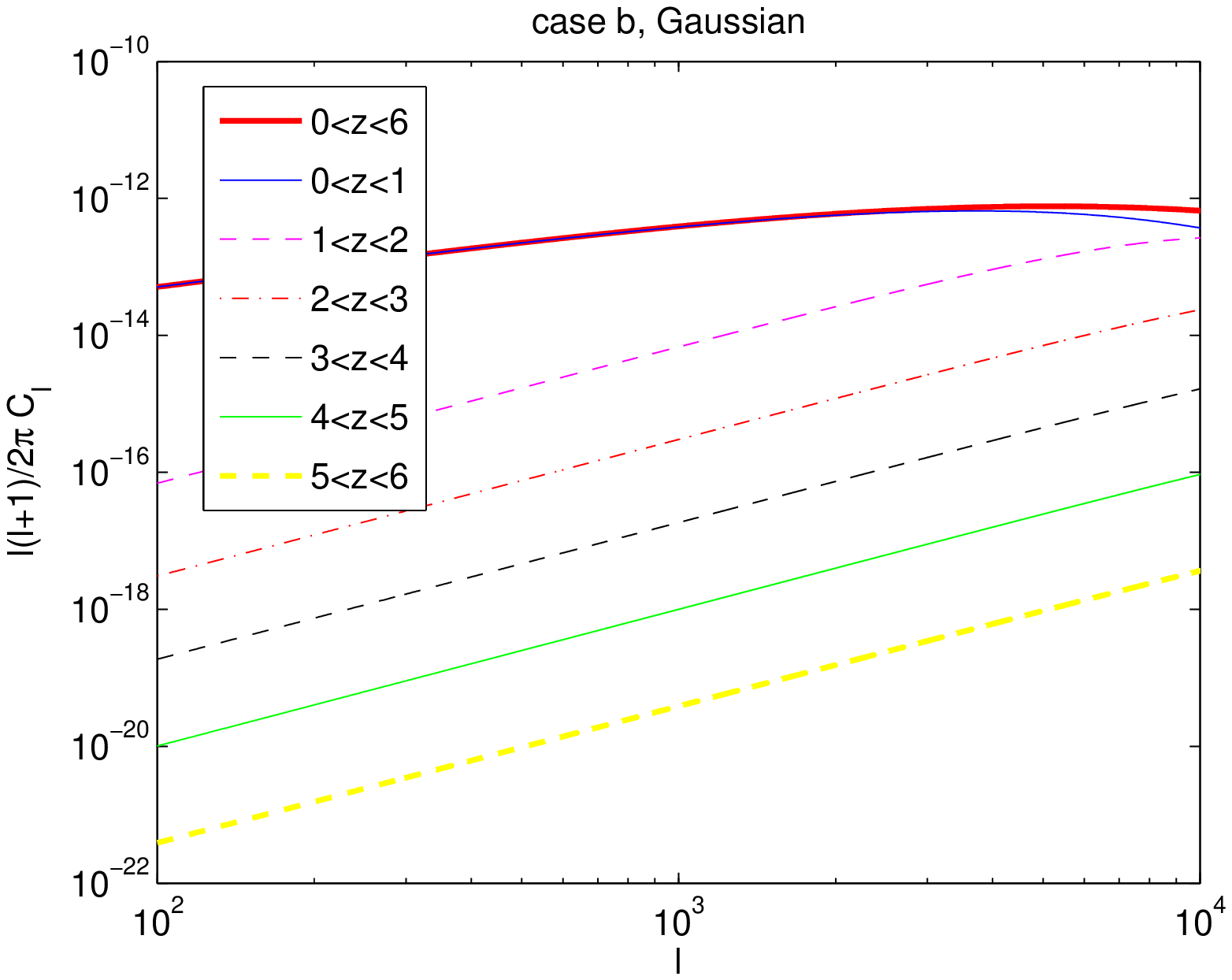, height=7.cm, width=8cm,clip=}
\epsfig{file=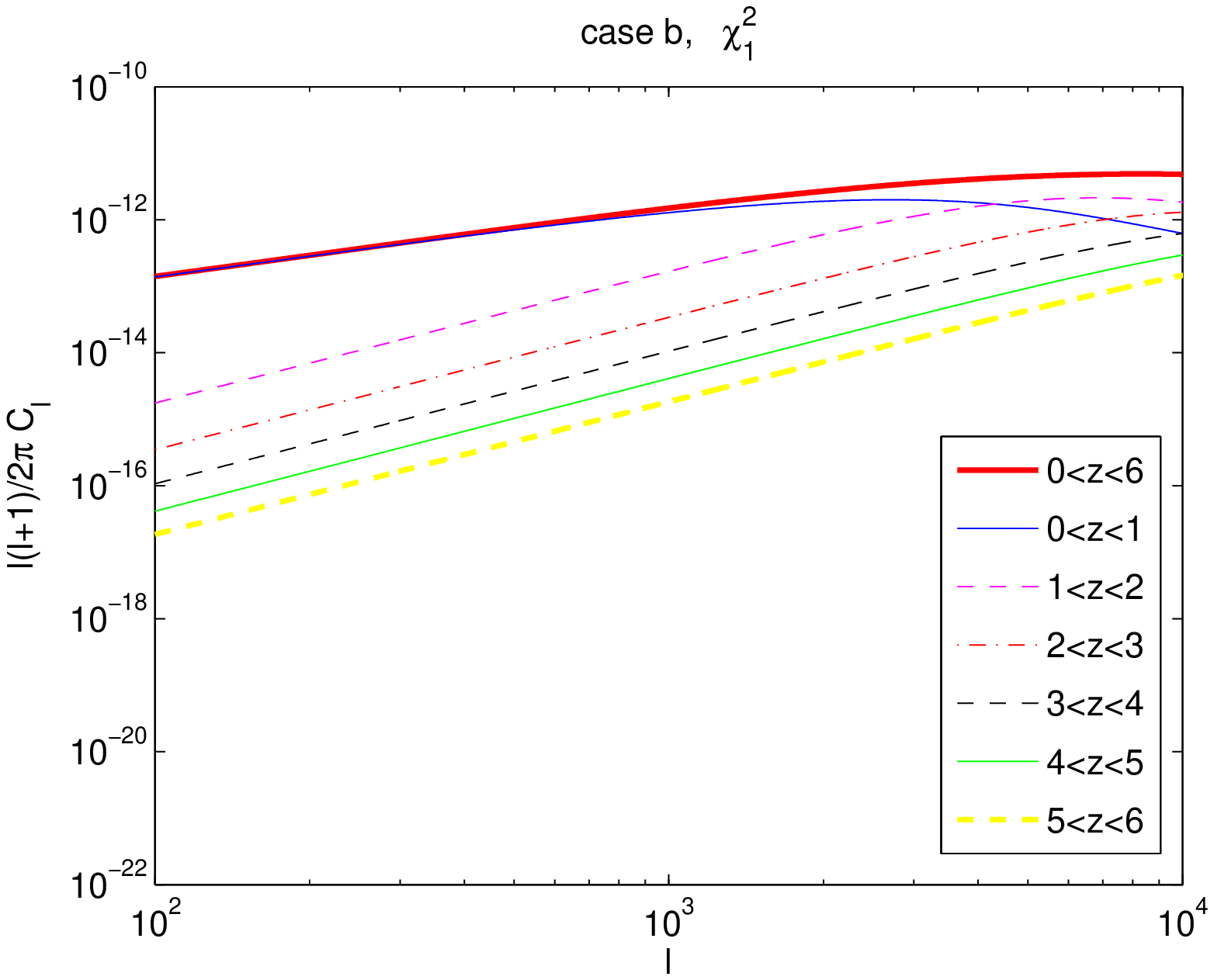, height=7.cm, width=8cm,clip=}
\caption{Angular power spectrum of the S-Z effect. Shown are levels of 
power contributed by clusters lying at redshift range of $0<z<6$, and at 6 
redshift intervals. Left and right-hand panels correspond to the Gaussian 
and $\chi^{2}_1$ models, respectively. Upper and lower panels relate 
to cases (a) and (b), respectively}.
\label{fig:clz}
\end{figure*}

\begin{figure*}
\centering
\epsfig{file=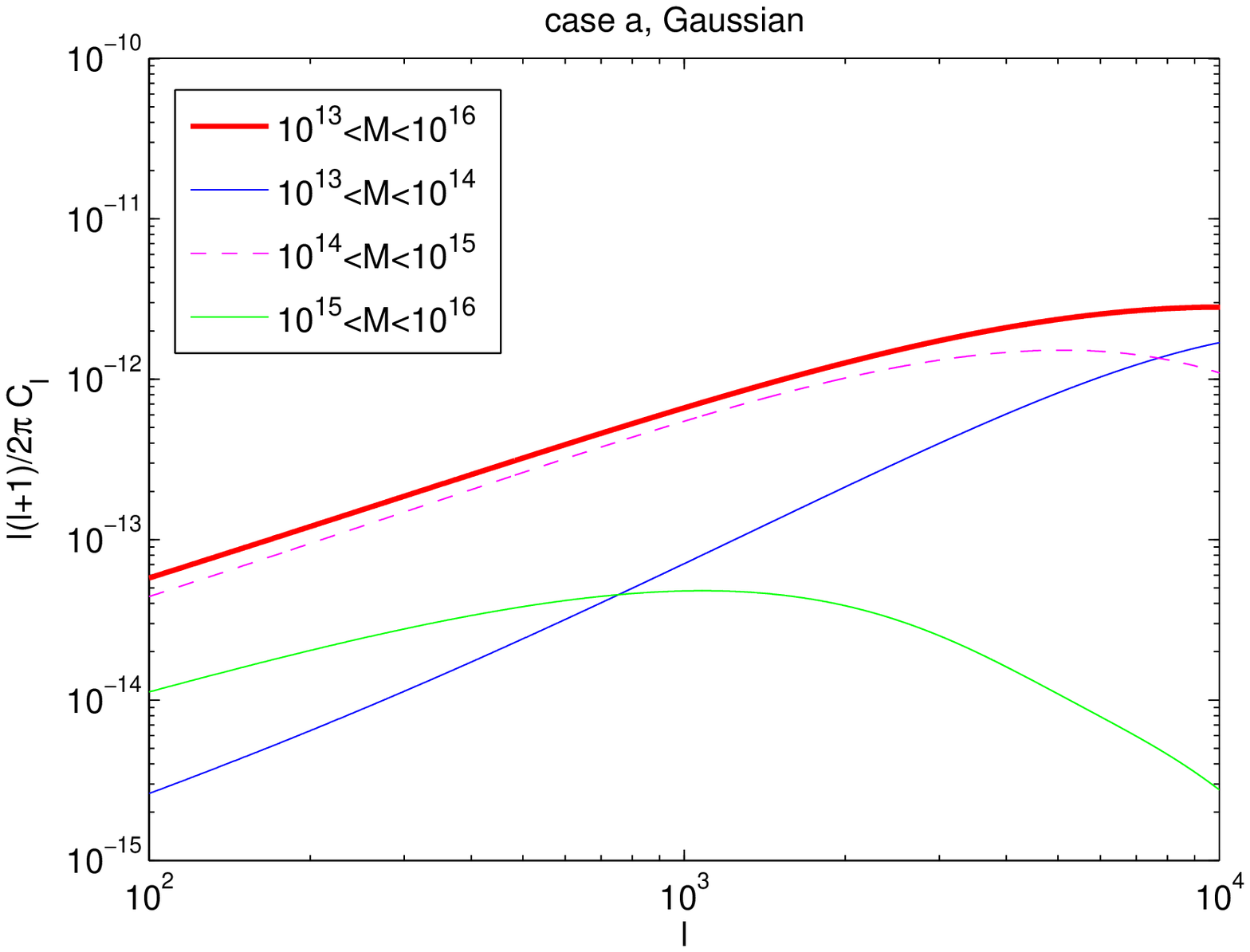, height=7cm, width=8cm,clip=}
\epsfig{file=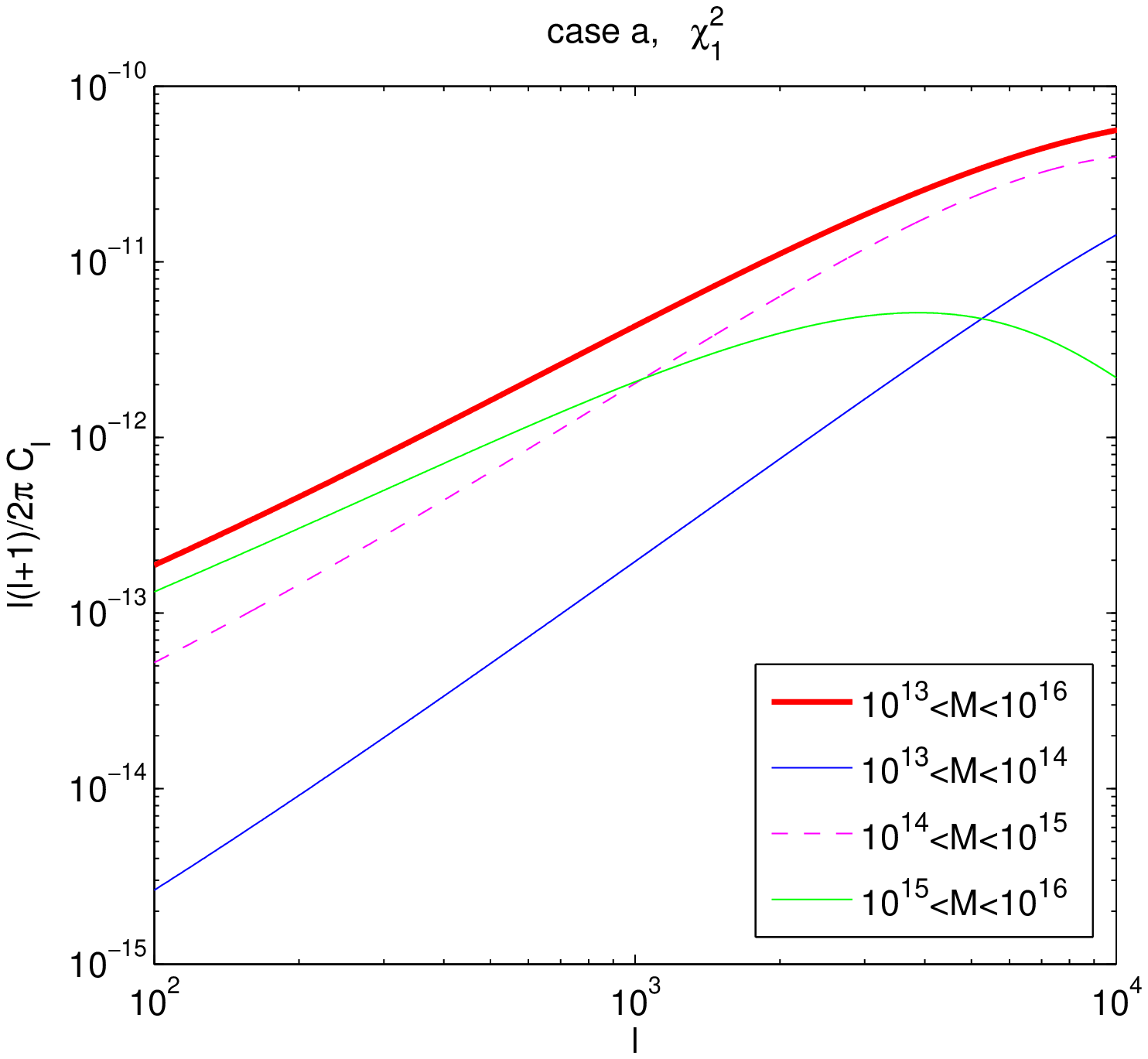, height=7cm, width=8cm,clip=}
\epsfig{file=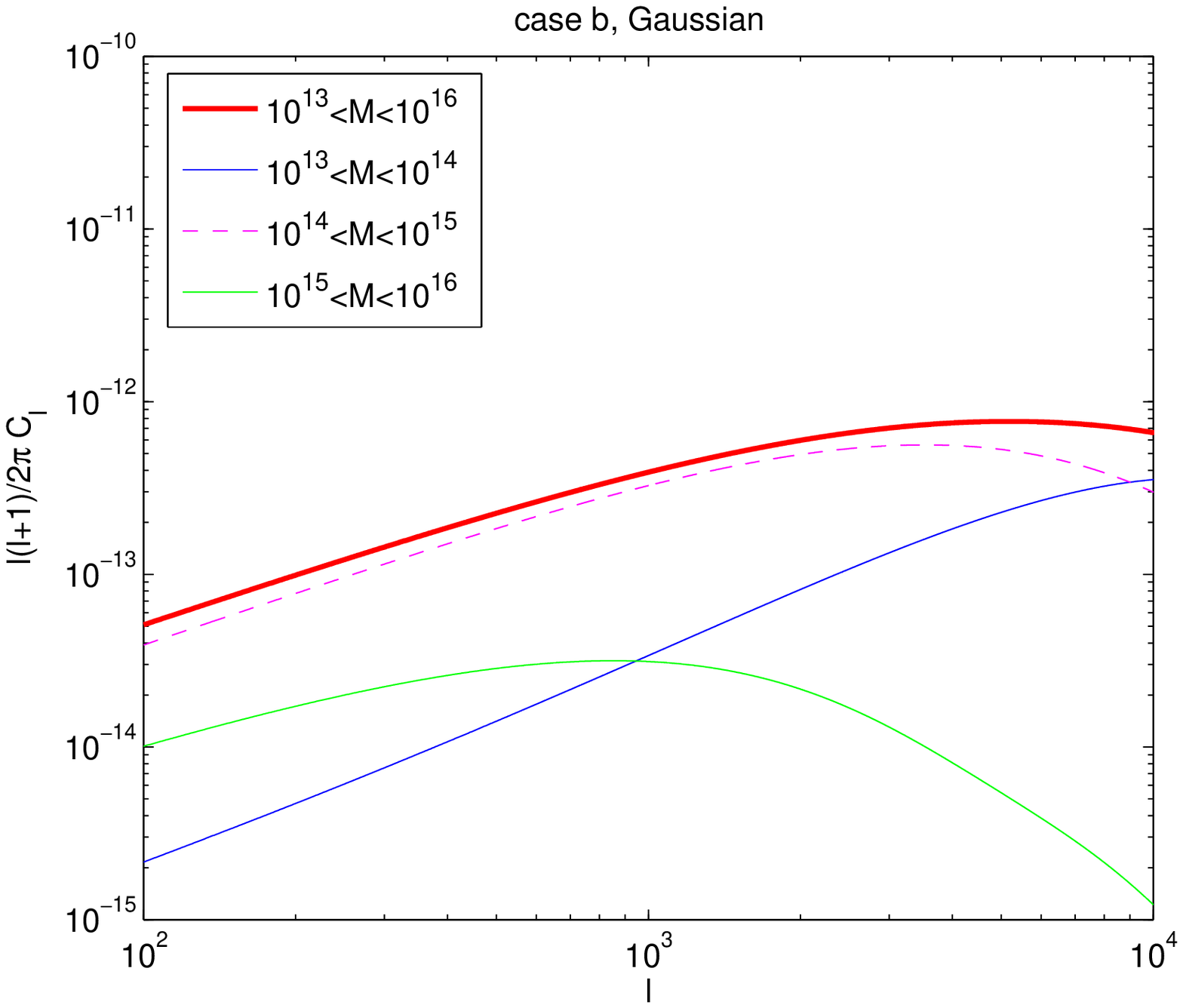, height=7cm, width=8cm,clip=}
\epsfig{file=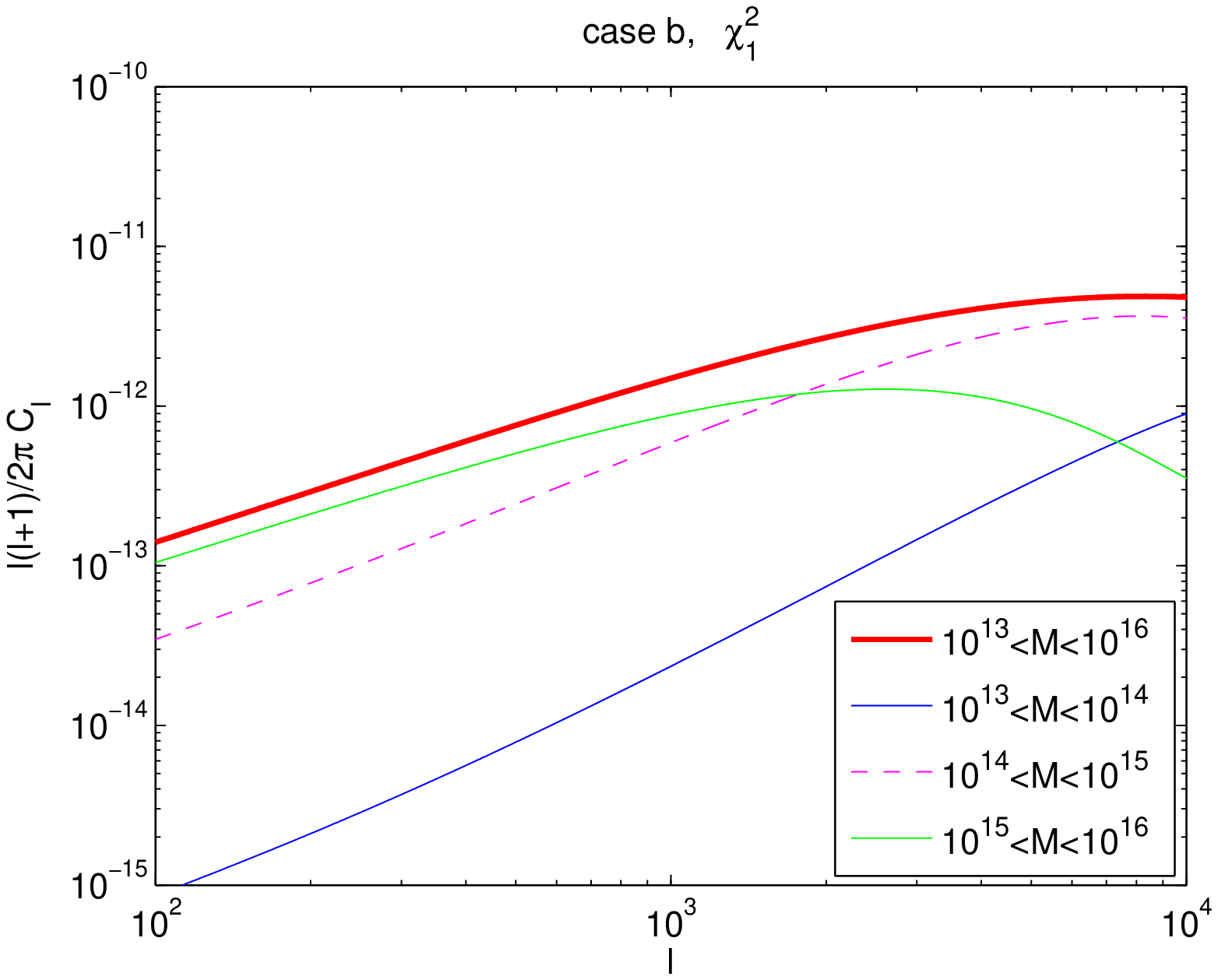, height=7cm, width=8cm,clip=}
\caption{The same as in the previous figure, but here power spectrum levels 
are calculated for 4 mass intervals, spanning the range 
$10^{13}h^{-1}M_{\odot}<M<10^{16}h^{-1}M_{\odot}$.}
\label{fig:clm}
\end{figure*}

\begin{figure*}
\noindent
\epsfig{file=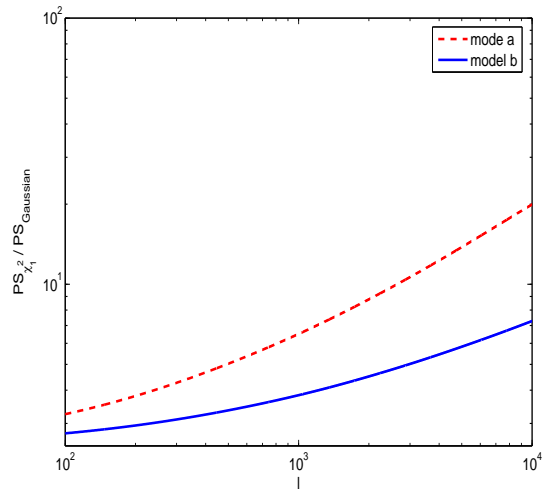, height=7cm, width=8cm,clip=}
\caption{A comparison of S-Z power spectrum levels in the $\chi^2_1$ and
Gaussian models for cases (a) and (b).}
\label{fig:clcomp}
\end{figure*}

\begin{figure*}
\centering
\epsfig{file=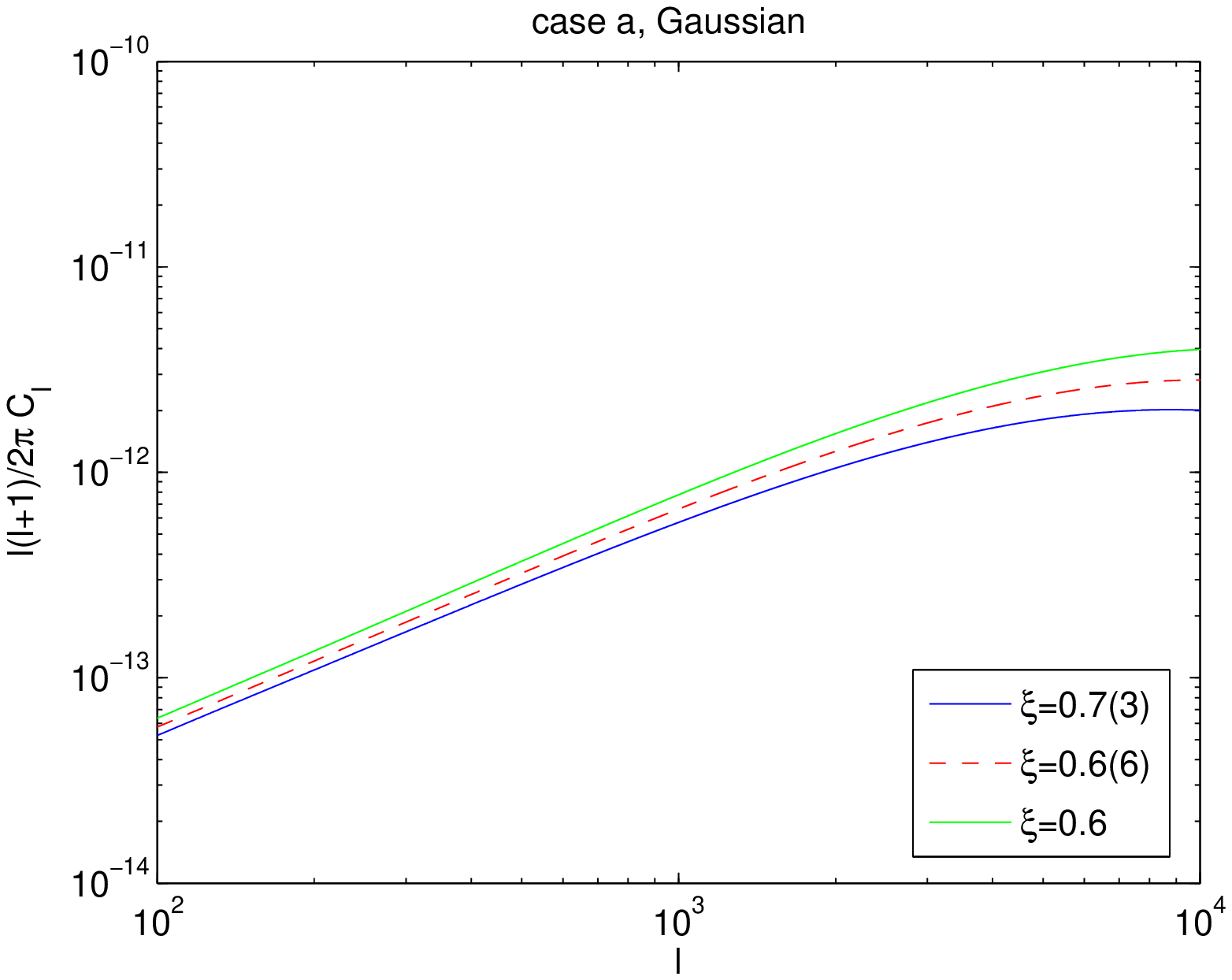, height=7.cm, width=8cm,clip=}
\epsfig{file=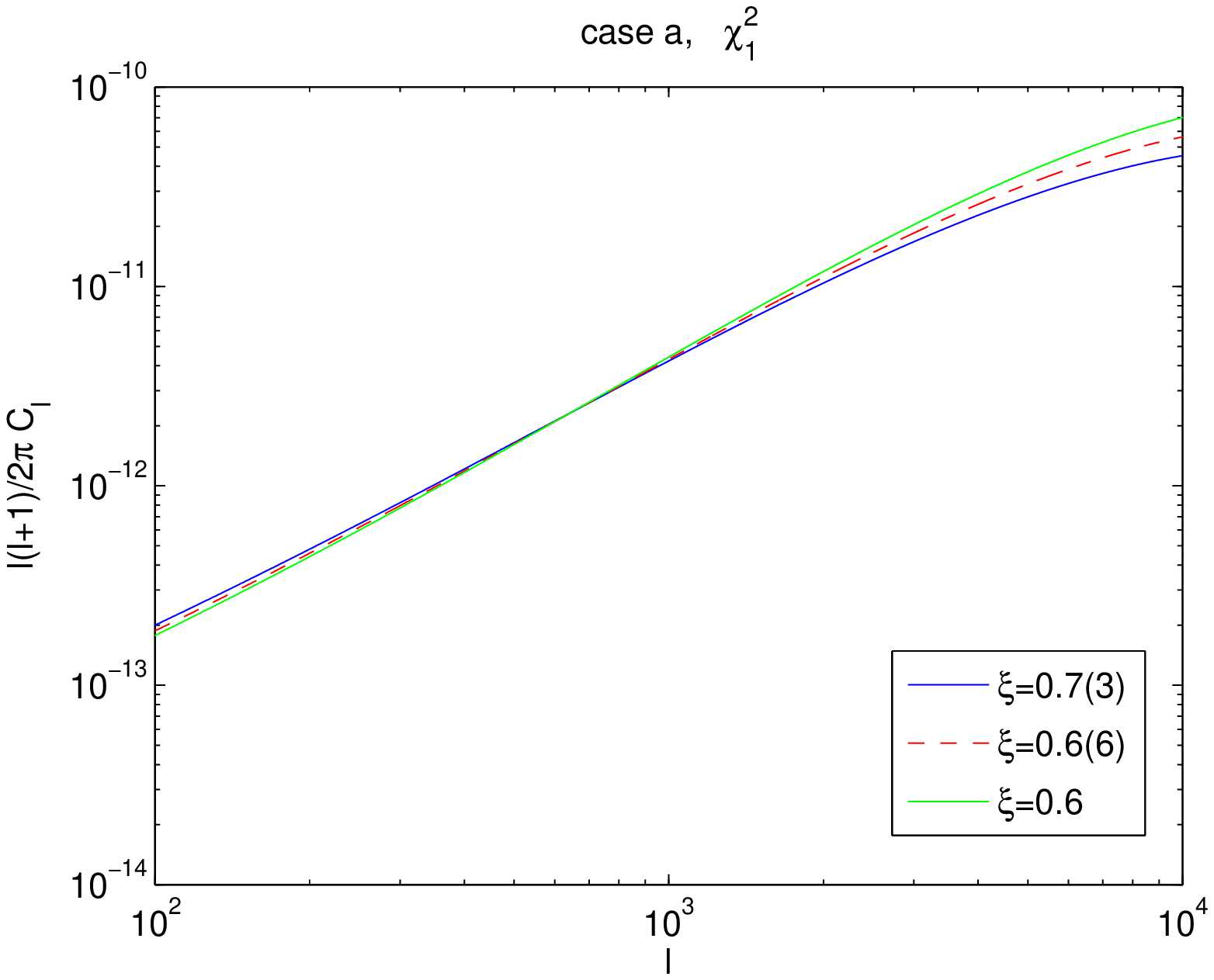, height=7.cm, width=8cm,clip=}
\epsfig{file=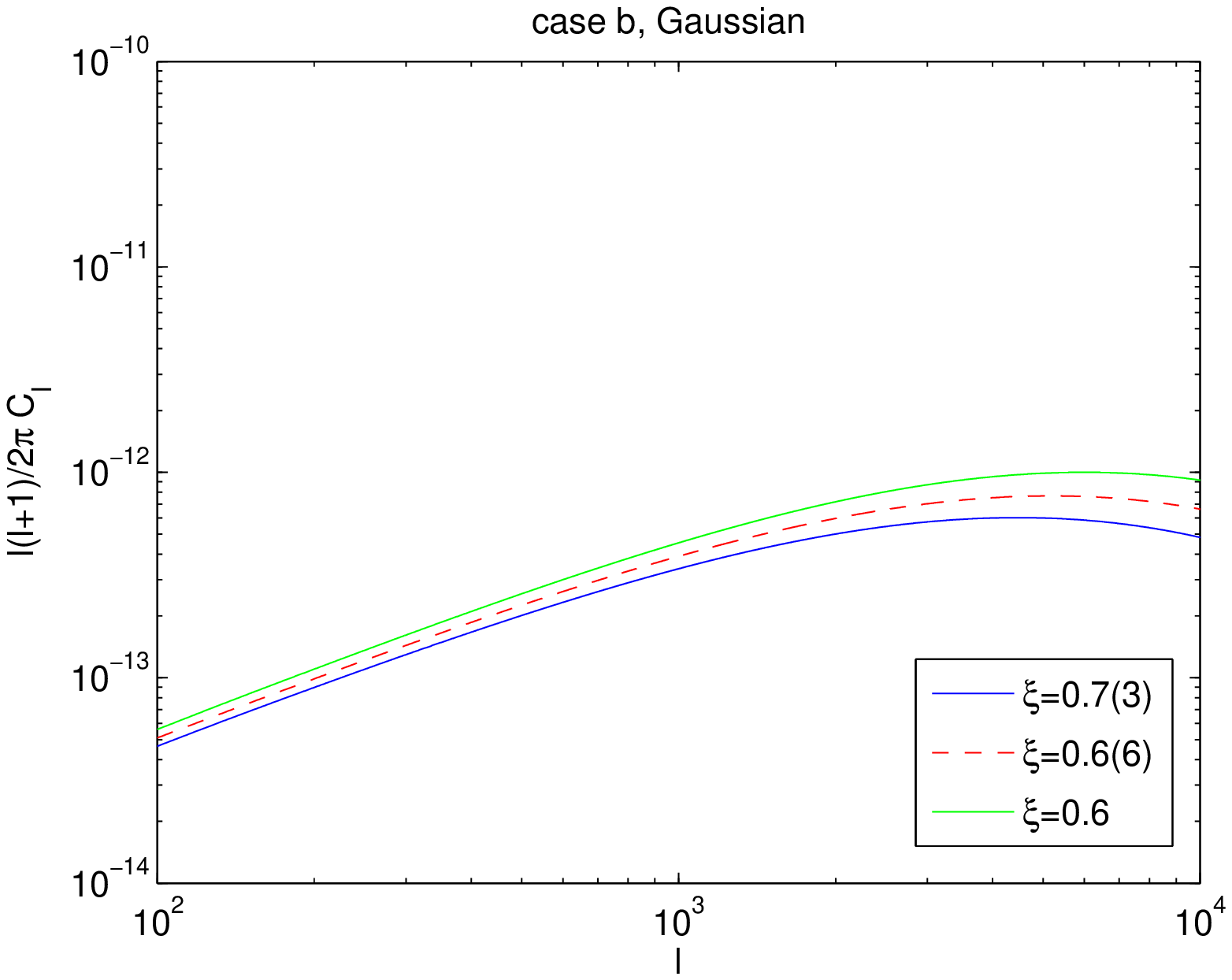, height=7.cm, width=8cm,clip=}
\epsfig{file=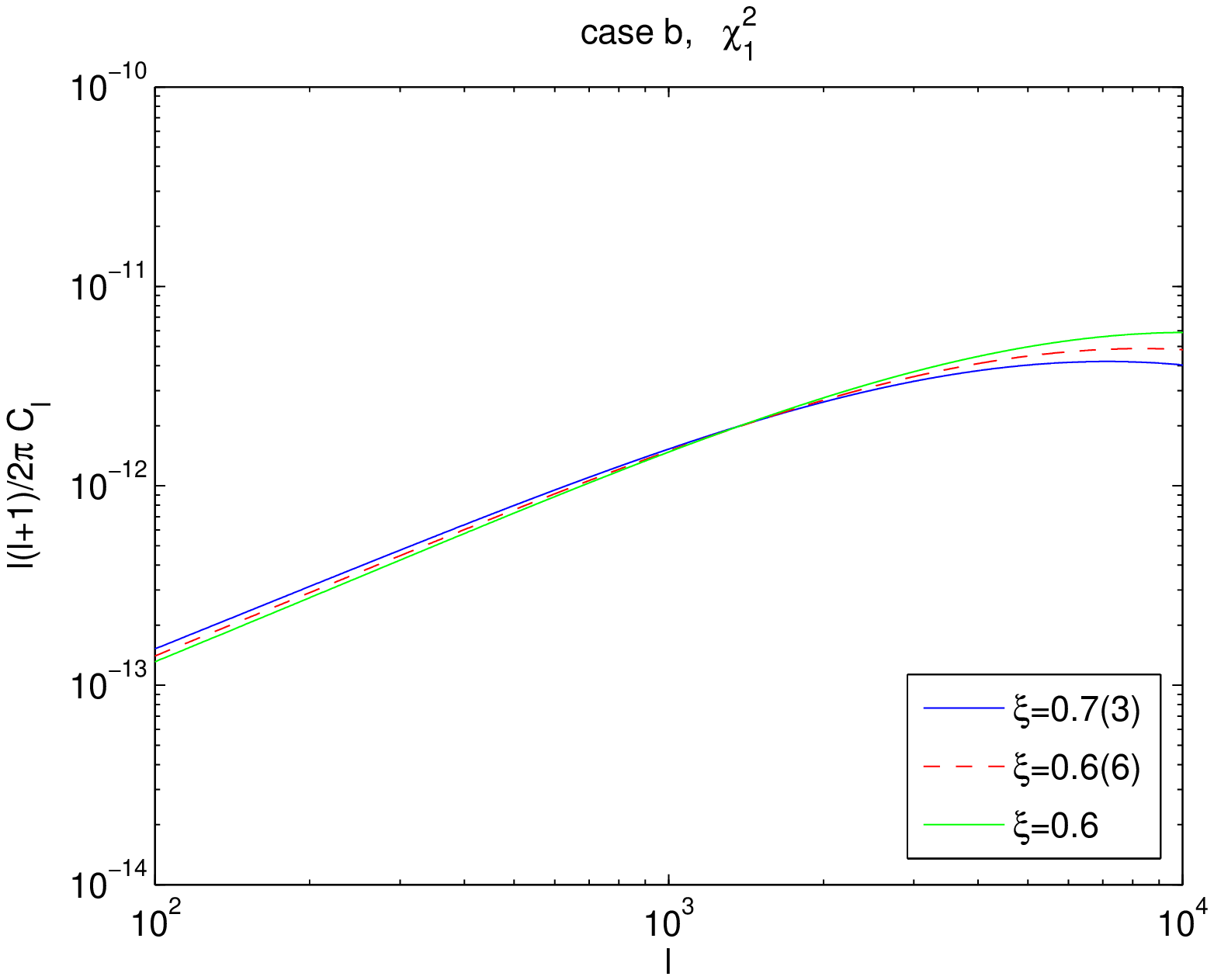, height=7.cm, width=8cm,clip=}
\caption{The impact of the temperature-mass relation on the S-Z power 
spectrum, $T\propto (M/M_{15})^{\xi}$. Shown are 3 cases: $\xi=0.7(3), 
\xi=0.6(6), \xi=0.6$. The arrangement of the panels is the same as in 
the previous figures.}
\label{fig:clxi}
\end{figure*}

\begin{figure*}
\centering
\epsfig{file=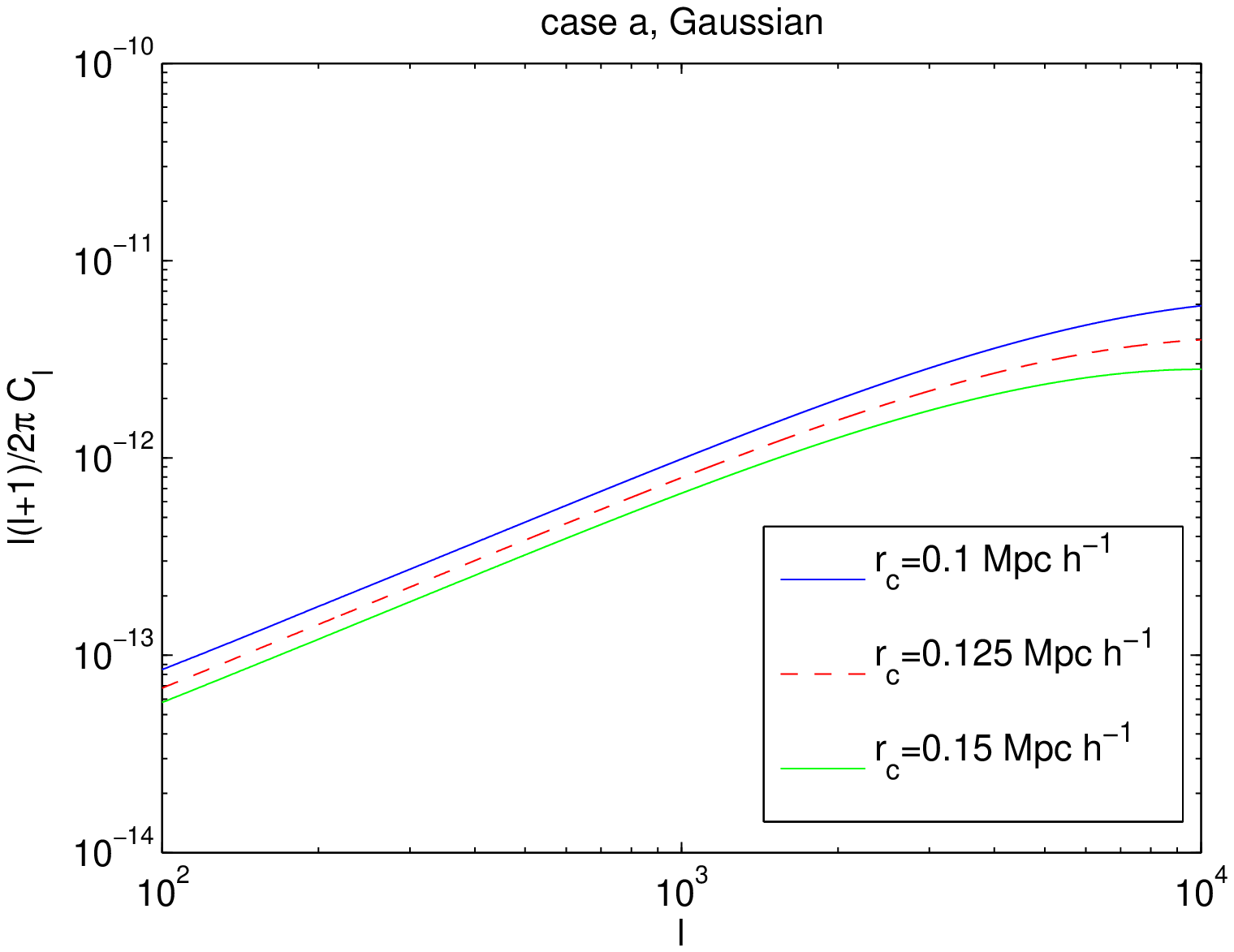, height=7.cm, width=8cm,clip=}
\epsfig{file=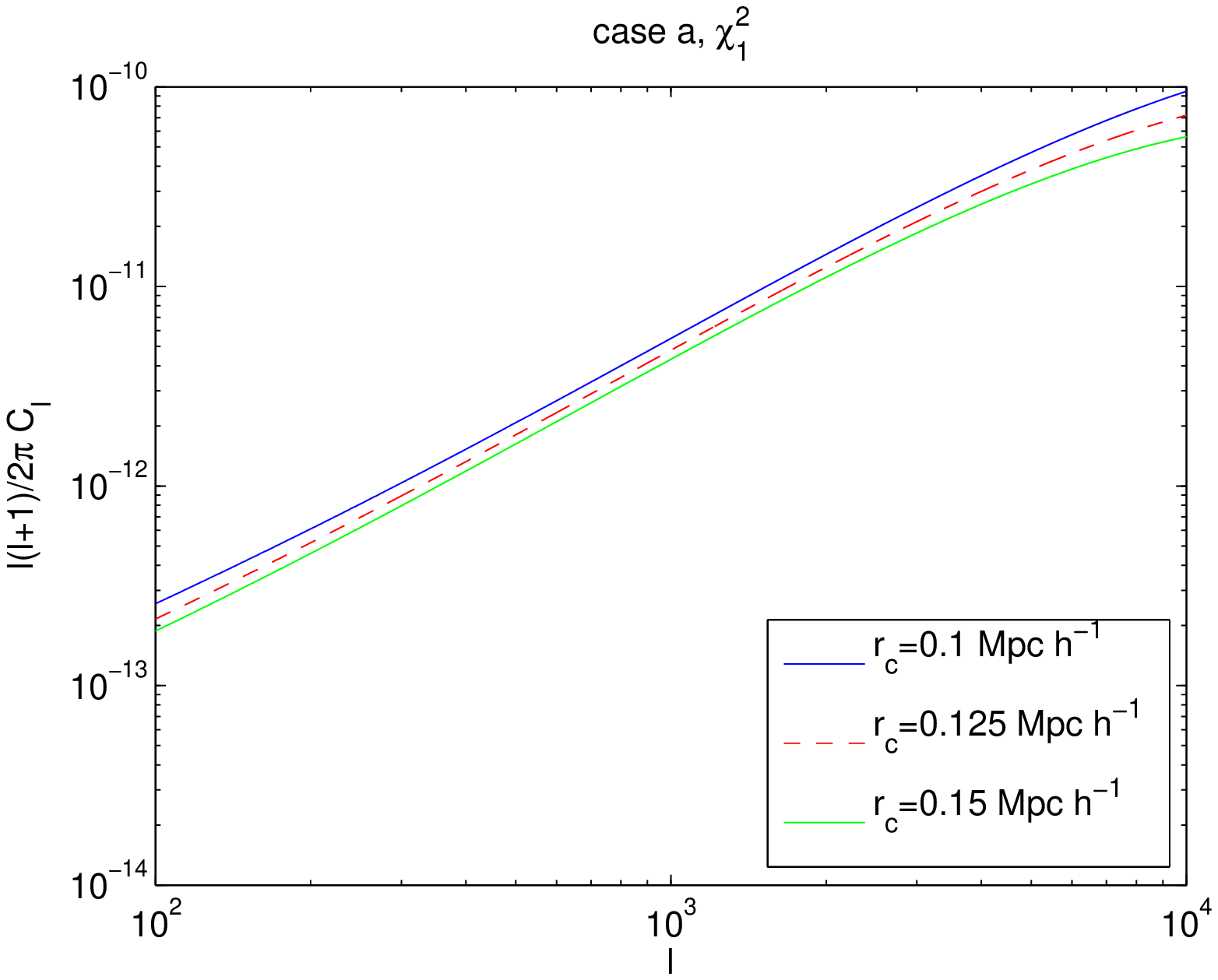, height=7.cm, width=8cm,clip=}
\epsfig{file=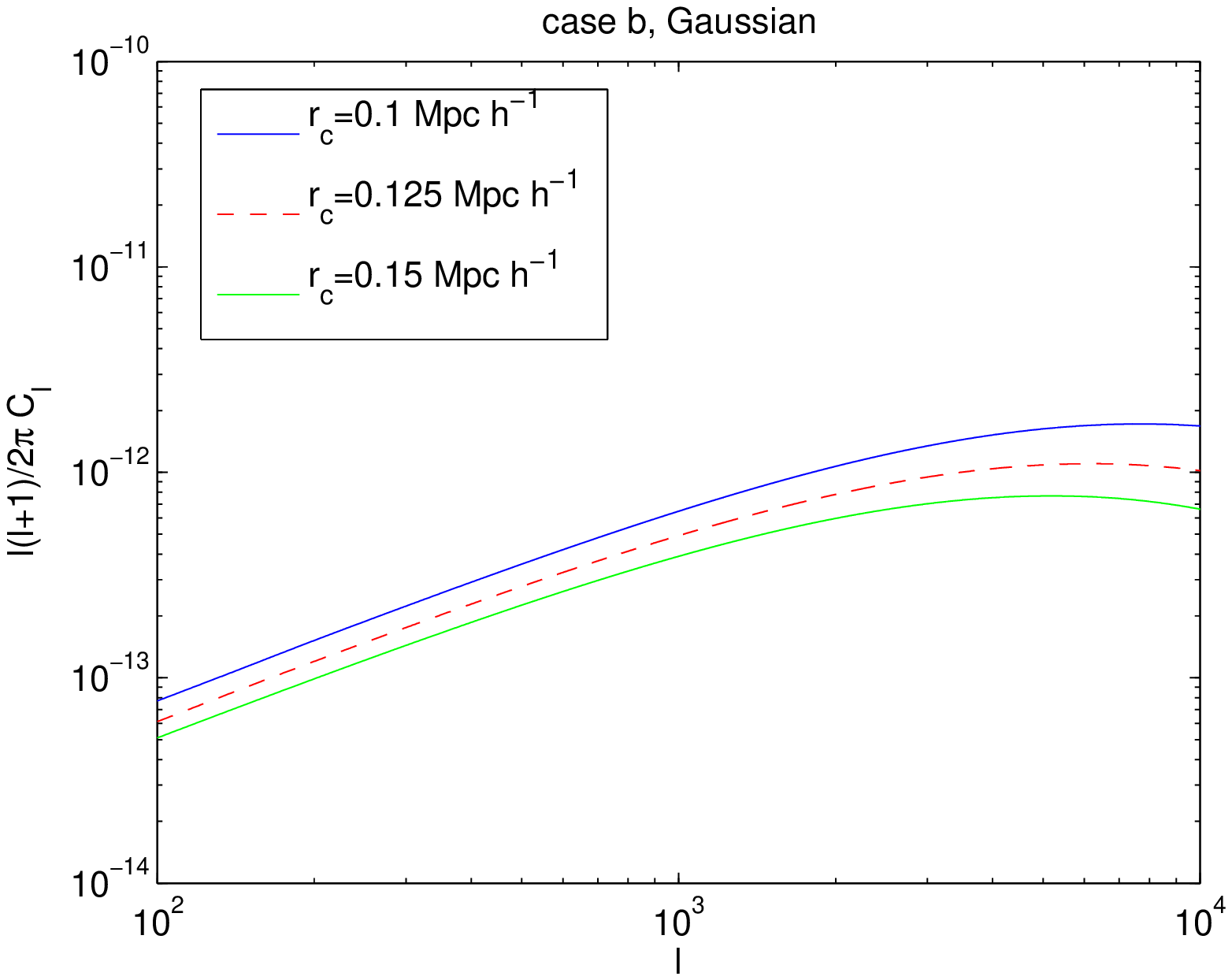, height=7.cm, width=8cm,clip=}
\epsfig{file=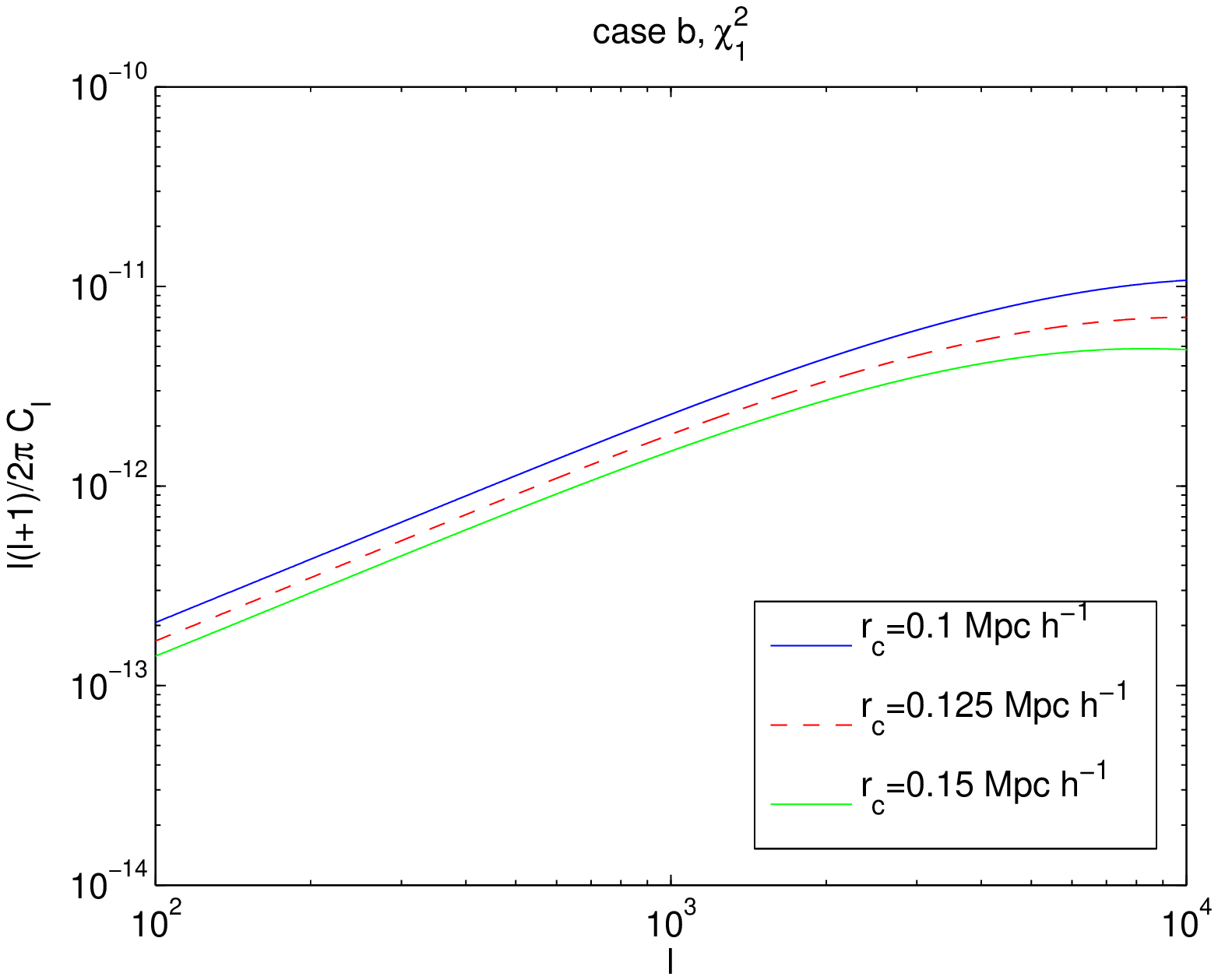, height=7.cm, width=8cm,clip=}
\caption{The impact of cluster core scaling on the S-Z power spectrum. 
Each of the three values considered 
($r_c=0.1, 0.125, 0.15\, Mpc\,h^{-1}$) denote the core radius of a 
cluster with mass $10^{15} M_{\odot}\,h^{-1}$ at redshift $z=0$.}
\label{fig:clrc}
\end{figure*}

The S-Z power spectrum calculated for cases (a) and (b) is plotted in 
Figs.~\ref{fig:clz} through~\ref{fig:clrc} (with the exception of 
Fig.~\ref{fig:clcomp}). 
In all plots the upper and lower panels correspond to cases (a) and (b), 
respectively, whereas the left- and right-hand panels pertain to the 
Gaussian and $\chi^{2}_1$ models, respectively. Partial contributions 
to the total power spectrum from redshift intervals of $\Delta z=1$ are 
shown in Fig.~\ref{fig:clz}. Fig.~\ref{fig:clm} depicts partial 
contributions from (logarithmic) mass intervals of $\log{\Delta M}=1$ to 
the total power spectrum. Power variation corresponding to a $10\%$ 
variance in the parameter $\alpha$, the scaling of cluster temperature 
with mass, is illustrated in Fig.~\ref{fig:clxi}. Finally, the impact 
of introducing a variance of $20\%$ in the cluster core scaling on the 
magnitude of the S-Z power spectrum is shown in Fig.~\ref{fig:clrc}. 

Referring first to the magnitude of the total power spectrum in 
Fig.~\ref{fig:clz}, corresponding to the redshift interval $0<z<6$, 
in both cases (a) and model (b) higher levels of power are clearly 
evident in the non-Gaussian model. Quantitatively, the maximum power 
levels attained in case (a) are $\sim 3\cdot 10^{-12}$ and 
$\sim 6\cdot 10^{-11}$ for the Gaussian and non-Gaussian models, 
respectively. Note that both peak at multipoles higher than $\ell>10,000$; 
in the Gaussian model this is due to the high gas densities of distant 
clusters implied by the constancy of the gas fraction, whereas in the 
non-Gaussian model the combination of high gas densities and the long 
tail at the high-mass cluster end is responsible for this result. Compared 
to case (a), distant clusters are cooler in case (b) by virtue of the 
redshift independence of their temperatures, resulting in lower power 
levels ($\sim 8^{-13}$ and $5\cdot 10^{-12}$) in the Gaussian 
and $\chi^{2}$ models, respectively. In the Gaussian model the power 
peaks at $\ell\sim 6000$, reflecting the lower contribution from cooler, 
more distant clusters. In the non-Gaussian model, a relatively high 
abundance of massive clusters at high redshifts (i.e, small angular 
scales) leads to sustained high power levels on these scales. It is 
also worth recalling that regardless of whether or not the temperature 
changes with redshift, it still increases with increasing mass, so 
even when $\psi=0$ a denser population of hot clusters is produced 
in the long tail of the $\chi^{2}$ PDF. 

The contribution of clusters lying at relatively high redshifts to the 
overall power is clearly seen in the $\chi^{2}$ model. In fact, in case 
(a) the contribution of clusters lying in the redshift range $0<z<1$ peaks
at $\ell\sim 3000$. However, the total power continues to rise owing to the
non-negligible power generated by clusters situated at $z>1$, particularly
so from redshift range $1<z<2$. In fact, at $\ell\sim 8000$ this 
contribution is surpassed by the next redshift range, $2<z<3$. In the 
Gaussian model the contribution from this range to the overall power is 
decidedly negligible. The same effect, although less pronounced, is 
evident in case (b); here distant clusters are cooler [than in the 
corresponding case (a)] by virtue of the $\psi=0$ scaling. Consequently, 
the partial contributions from redshift ranges $1<z<3$ is still 
discernible, but to a lesser degree than in case (a). 

Fig.~\ref{fig:clm} illustrates the distribution of S-Z power in four 
mass ranges. Most striking is the relative contribution to power from 
clusters lying in the $10^{15}h^{-1}M_{\odot}<M<10^{16}h^{-1}M_{\odot}$ 
mass range. Whereas it is practically negligible at all multipoles in the 
Gaussian model, this is not the case in the non-Gaussian model; in fact, 
these massive clusters dominate the total power up to $\ell\sim 1000$ and 
$\ell\sim 2000$ in cases (a) and (b), respectively. At higher multipoles 
the $10^{14}h^{-1}M_{\odot}<M<10^{15}h^{-1}M_{\odot}$ range dominates. In 
the Gaussian model low-mass clusters of 
$10^{13}h^{-1}M_{\odot}<M<10^{14}h^{-1}M_{\odot}$ prevail only at the 
highest multipoles. These clusters are obviously the most abundant objects 
predicted by a mass function based on a primordial density fluctuation 
field of Gaussian nature, and  - in light of their relative small sizes 
(and low temperatures) - thus must lie at a relative proximity to the 
observer; more distant clusters of similar masses would not produce 
sufficiently strong S-Z signal to be detected. Otherwise, in the 
Gaussian case, the lion's share of power originates in clusters in the 
mass range $10^{14}h^{-1}M_{\odot}<M<10^{15}h^{-1}M_{\odot}$. 

Fig.~\ref{fig:clz} and~\ref{fig:clm} explicitly demonstrate the 
significantly higher contribution of high-redshift - and therefore 
hotter, in particular in case (a) - clusters in the $\chi^2_1$ model. 
This is manifested in higher power magnitude and at an intensification 
at higher multipoles corresponding to distant, low angular-size clusters.
To further contrast the predictions of the two models, the ratio of total 
level of the S-Z power in the $\chi^2_1$ model to that in the Gaussian 
model for cases (a) and (b) is shown in Fig.~\ref{fig:clcomp}. For 
$\ell =10^3 - 10^4$, this ratio increases from $7$ to $20$ in case (a), 
and from $\sim 4$ to $\sim 7$ in case (b).

The effect of changing $\alpha$, the scaling of the cluster temperature
with mass, is shown in Fig.~\ref{fig:clxi}. Since 
$T\propto (M/M_{15})^{\alpha}$, two different behaviours are expected for 
$M<M_{15}$ and $M>M_{15}$. For $M<M_{15}$, increasing $\alpha$ with 
respect to the conventional value of $2/3$ results in cooler clusters, 
whereas if $M>M_{15}$, higher temperatures will be attained. Lowering 
$\alpha$ results in the opposite effect, i.e., warmer low-mass clusters 
and cooler high-mass clusters. These changes affect the power spectrum 
for the Gaussian and non-Gaussian models in slightly different ways; we 
have already seen that in the Gaussian model the largest contribution to 
the power spectrum comes from clusters with masses 
$10^{14}h^{-1}M_{\odot}<M<10^{15}h^{-1}M_{\odot}$. That is, an outcome of
increasing $\alpha$ is that most of the contributing clusters undergo 
cooling, and the level of power decreases. Obviously, lowering $\alpha$ 
has the opposite effect. This is clearly seen in the upper- and 
lower-left panels of Fig~\ref{fig:clxi}. It can also be seen 
(particularly so in the lower-left panel) that the curves peak at 
decreasing multipoles with increasing $\alpha$. This has to do with the 
fact that the contribution of these relatively low-mass clusters to the 
overall power levels decreases with increasing $\alpha$, so that power 
is suppressed from high multipoles, resulting in a leftward shift of the 
peaks. In the $\chi^2_1$ model the results are somewhat different. Here 
the contribution to the power from high-mass clusters 
($10^{15}h^{-1}M_{\odot}<M<10^{16}h^{-1}M_{\odot}$) dominates at 
$\ell\lesssim 1000$ (case a) and $\ell\lesssim 2000$ (case b), whereas 
the corresponding contribution at higher multipoles originates in 
clusters lying within the mass range 
$10^{14}h^{-1}M_{\odot}<M<10^{15}h^{-1}M_{\odot}$. In other words, an 
increment of power with increasing $\alpha$ is expected at lower 
multipoles, and a decrease of power with increasing $\alpha$ at higher 
multipoles. This is shown in the upper- and lower-right panels of 
Fig.~\ref{fig:clxi}. 

We also investigate the influence of $r_c$ on the 
power spectrum levels. The results are illustrated in 
Fig.~\ref{fig:clrc}, and are rather obvious. First, lowering or 
increasing the core radius changes the effective cluster size, and 
consequently, the multipole at which the power peaks. This can be easily 
seen in the lower-left panel of Fig.~\ref{fig:clrc}. In the remaining 
panels it is somewhat harder to notice the effect, owing to the fact 
that the peaks fall beyond $\ell=10000$. Also, in order to keep the 
total gas mass constant, lower and higher core sizes necessitate higher 
and lower central gas densities, respectively. This results in higher 
and lower power levels for lower and higher core-sized clusters, 
respectively, as is apparent from the fact that the power spectrum 
scales as the central gas density squared 
(equations~\ref{eq:clmb} and~\ref{eq:ythet}). 

\begin{figure*}
\centering
\epsfig{file=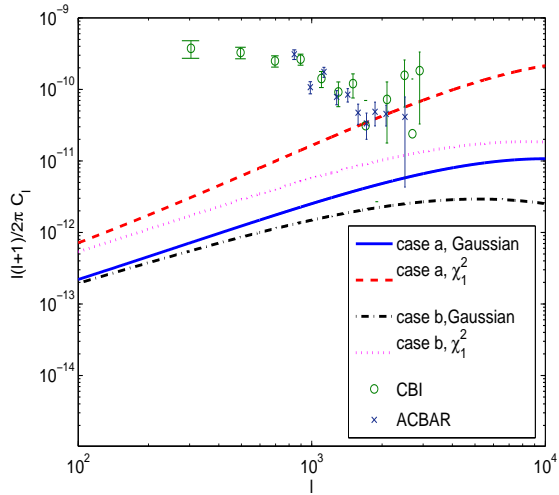, height=7.cm, width=8cm,clip=}
\caption{Total angular power spectrum of the S-Z effect as calculated above, 
together with results from the \textit{CBI} and \textit{ACBAR} experiments.}
\label{fig:cbiacb}
\end{figure*}

It is useful to mention some recent studies of S-Z power spectrum calculations 
and compare with our results. Case (a) of 
the Gaussian model is the most adequate for comparison, since past work has 
focused on Gaussian density fields, and whether analytically or inferred from 
hydrodynamical simulations, the IC gas temperature has been assumed (or 
observed) to increase with redshift (at least to a certain degree), in contrast 
with the constant temperature scenario, case (b). Da Silva et al. (2001) 
employed 
hydrodynamical simulations to find power levels of $5-9\cdot 10^{12}$ peaking 
at $\ell\sim 7000$, depending on whether or not non-adiabatic processes such as 
preheating and radiative cooling were included. Their selection of cosmological 
parameters is essentially identical to ours, but the calculations were carried 
out in the Rayleigh-Jeans region, so a comparison between their results and ours
requires an introduction of a multiplicative factor of 4 to our power levels.
Consequently, the magnitude of the total power for case (a) in the Gaussian model
peaks at multipole $\ell\sim 10,000$, at a value of $\sim 1.2\cdot 10^{-11}$.
Bond et al. (2005) have quantified the S-Z power spectrum using both
analytic calculations and results of hydrodynamical simulations. 
The simulations yield peak power levels of $1\sim 4\cdot 10^{-11}$ at 
$\ell\sim 10,000$. 
Their analytic calculation yields similar power levels, but at lower 
peak multipoles, $\ell\sim 5000-6000$. 
Note that they calculated S-Z power at the \textit{CBI} frequency, $\nu=31\,GHz$. 
Rescaling our results to this frequency gives a power level of 
$\sim 1.1\cdot 10^{-11}$. Refregier \& Teyssier (2002) have also 
used hydrodynamical simulations and an analytic approach to calculate the 
S-Z power spectrum. They found that both methods yield approximately the same power 
levels (peaking at $\ell\sim 8000$ with magnitude $\gtrsim 10^{-11}$), particularly 
so when a cluster mass range $5\cdot 10^{10}<M<8\cdot 10^{14}\,h^{-1}\,M_{\odot}$ is 
employed in the analytical calculation. The calculations were carried out in 
the R-J region. Springel, White \& Hernquist (2001) 
employed hydrodynamical simulations, neglecting non-adiabatic processes, to infer power 
levels of $2-4\cdot 10^{-11}$ at the peak, $\ell\sim 10,000$. 
In fact, the latter paper presents a detailed comparison of S-Z power 
spectra published in the literature showing a large scatter 
(spanning over an order of magnitude), which can largely 
be attributed to uncertainties in the theoretical modeling of 
IC gas and the techniques implemented in hydrodynamical calculations. 
To complete this section, we combine, in a single plot, 
the total power spectrum levels of the S-Z effect in the Gaussian and 
$\chi^2_1$ models, cases (a) and (b), with results from
the \textit{CBI} and \textit{ACBAR} experiments. 
The analytic results were rescaled so as to correspond to an observation 
frequency of 31 GHz, corresponding to the \textit{CBI} instrument. As can be 
seen in Fig.~(\ref{fig:cbiacb}), 
it seems difficult to reconcile the Gaussian results with 
the data, whereas case (a) in the $\chi^2_1$ model provides a reasonable match to the 
observed power excess at $\ell\gtrsim 2000$. These results agree with those 
found by Mathis et al. (2004). 

\subsection{Number counts}
\begin{figure*}
\centering
\epsfig{file=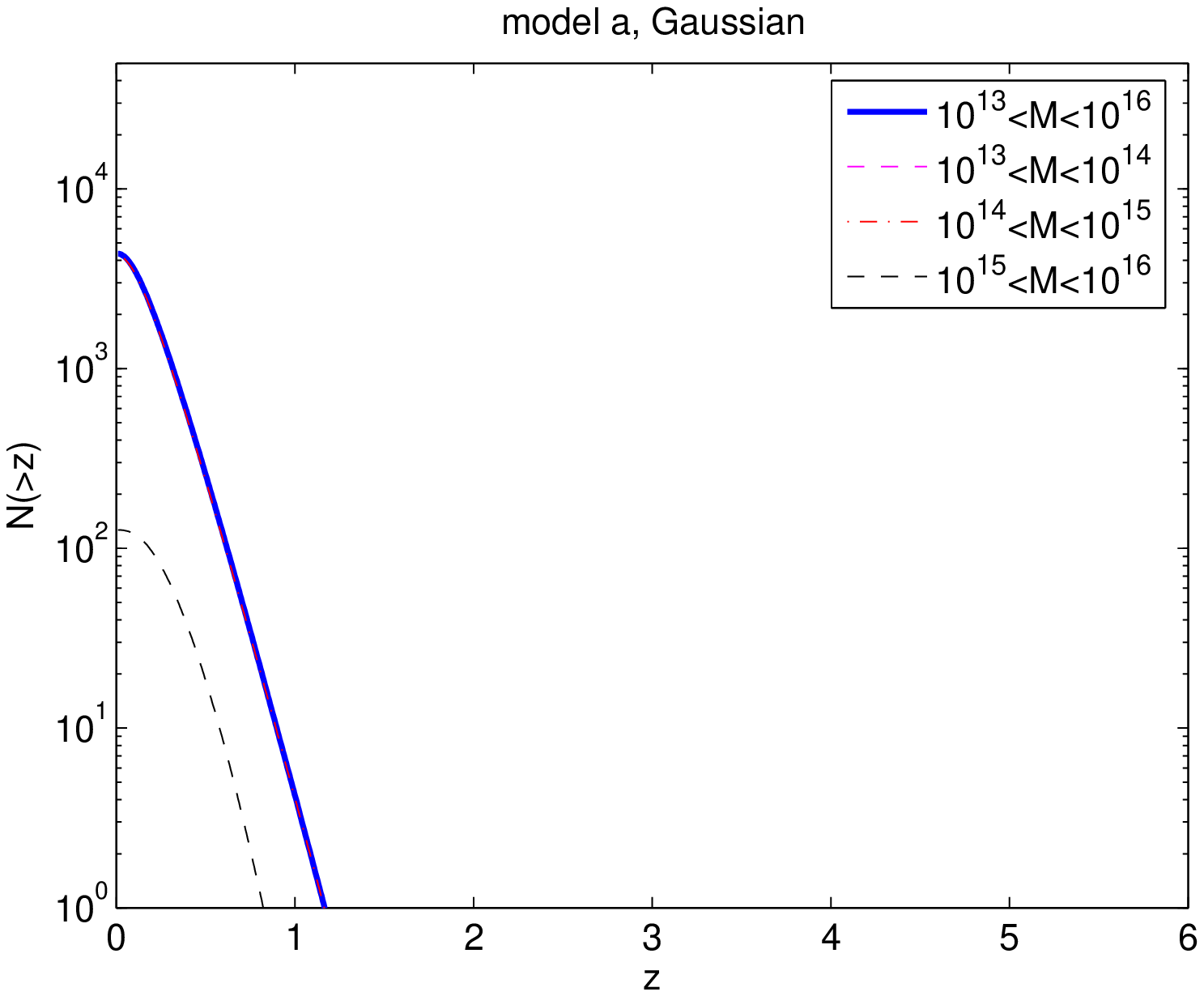, height=7.cm, width=8cm,clip=}
\epsfig{file=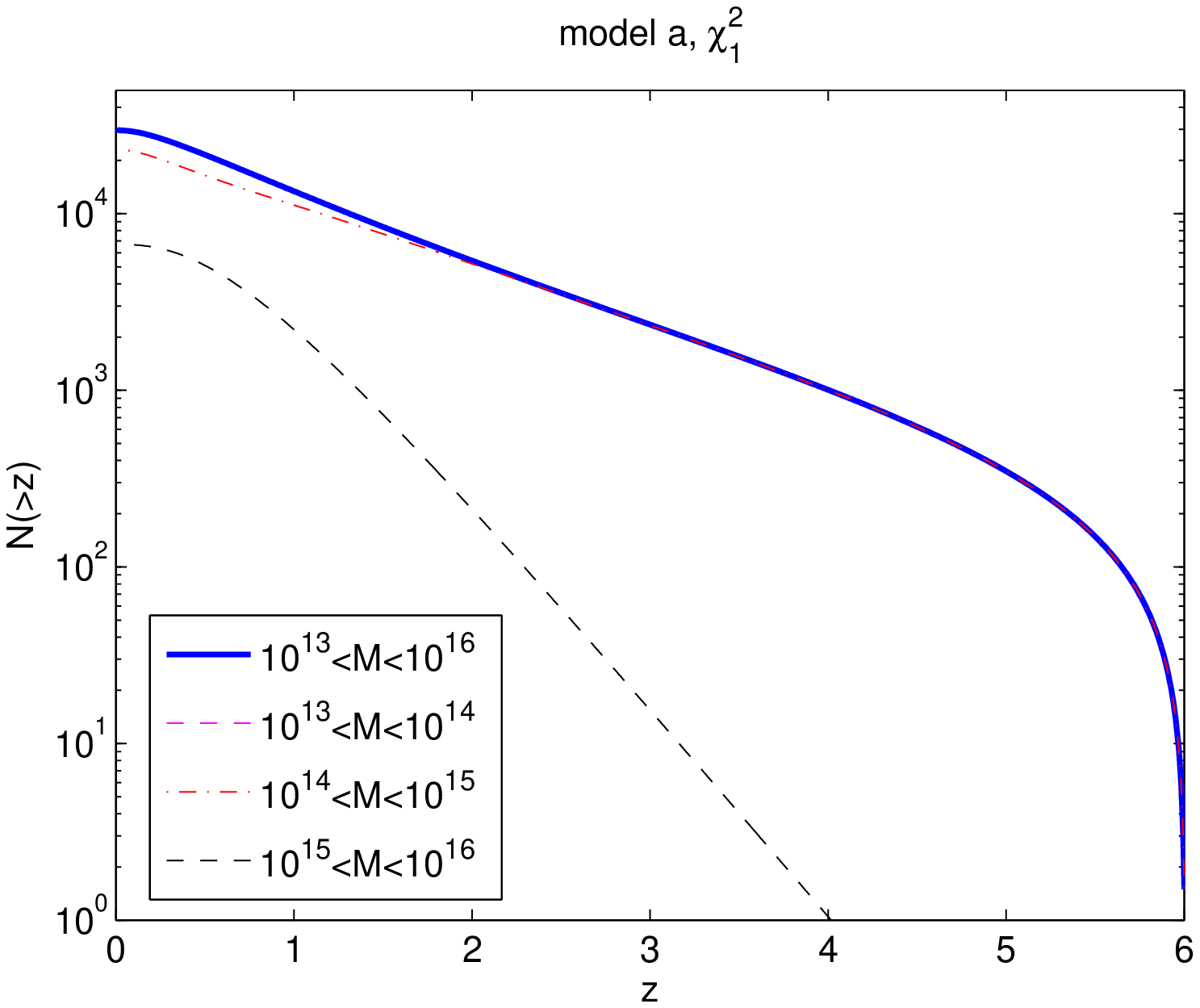, height=7.cm, width=8cm,clip=}
\epsfig{file=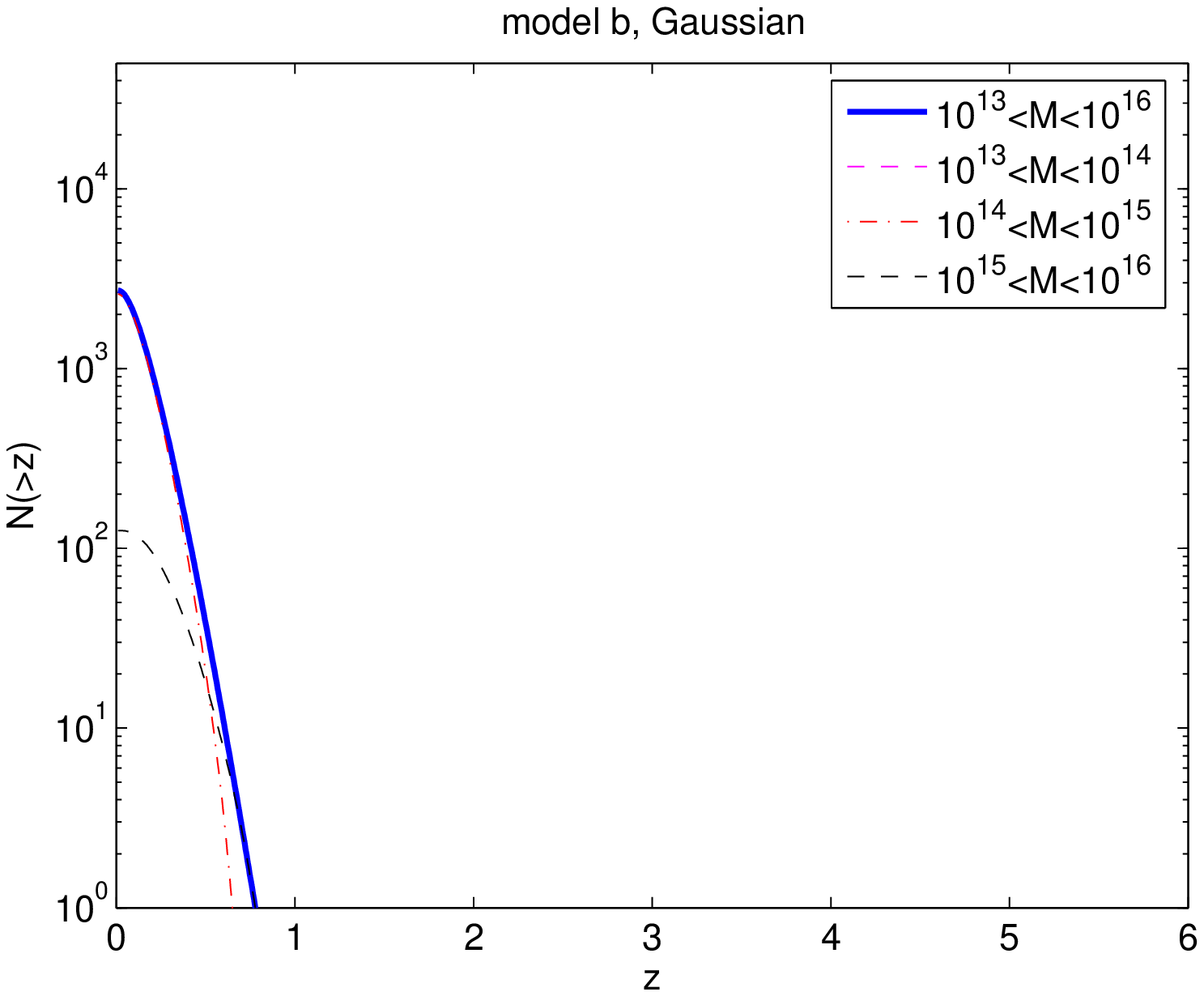, height=7.cm, width=8cm,clip=}
\epsfig{file=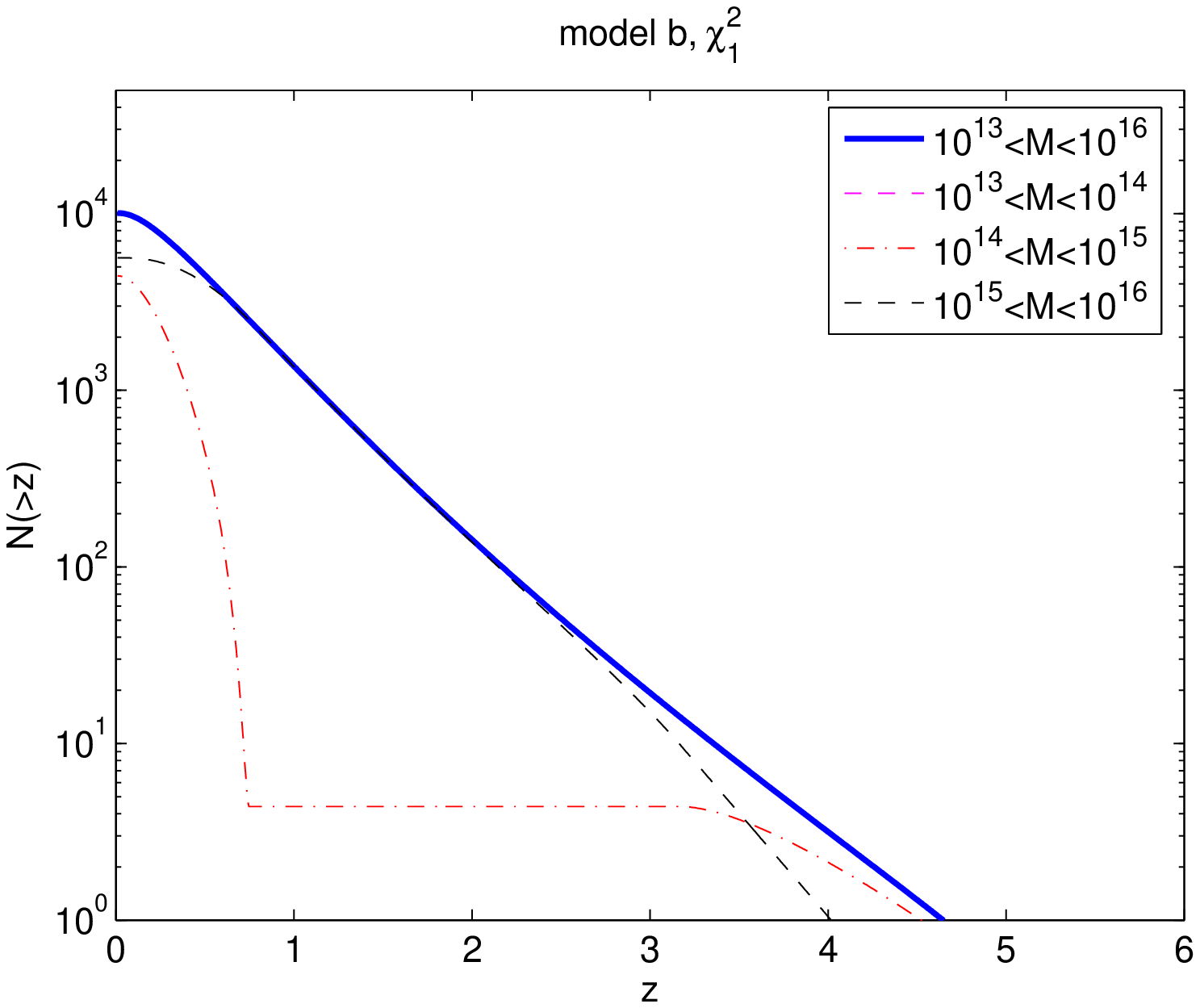, height=7.cm, width=8cm,clip=}
\caption{Cumulative S-Z counts as a function of redshift; shown are 
contributions from 4 mass intervals. Note that the contribution from 
the lowest mass interval ($10^{13}<M<10^{14}$ solar masses) vanishes 
in all cases considered.}
\label{fig:nctz}
\end{figure*}

The qualitative considerations pertaining to the power spectrum that were 
presented in the previous section apply also to S-Z number counts: Higher 
(cumulative) counts are expected in the non-Gaussian case, and those are 
likely to be distributed differently in redshift space with respect to the 
Gaussian case, reflecting the higher population of massive clusters at 
high redshifts, which satisfy the detection flux limit. 
Fig.~\ref{fig:nctz} depicts the redshift distribution of cumulative 
S-Z number counts. The arrangement of plots according to model is the 
same as in Fig.~\ref{fig:clz} through~(\ref{fig:clrc}). In none of 
the models clusters with masses lower than $10^{14}h^{-1}M_{\odot}$ 
generate fluxes that exceed the adopted detection limit, $30\,mJy$. 

The most noticeable results displayed in the figure are: In the Gaussian 
model the largest contribution to the cumulative number counts (at $z=0$) 
comes from clusters lying in the mass range 
$10^{14}h^{-1}M_{\odot}<M<10^{15}h^{-1}M_{\odot}$. Actually, the 
cumulative number counts at $z=0$ in this mass range are higher by at 
least an order of magnitude than those contributed by clusters in the 
mass range $10^{15}h^{-1}M_{\odot}<M<10^{16}h^{-1}M_{\odot}$. 
Specifically, in case (a) the curves depicting the total counts and 
those contributed by clusters with masses in the range 
$10^{14}h^{-1}M_{\odot}<M<10^{15}h^{-1}M_{\odot}$ are virtually 
identical. In case (b), however, at redshifts $z\gtrsim 0.5$ counts are 
dominated by clusters with higher masses, 
$10^{15}h^{-1}M_{\odot}<M<10^{16}h^{-1}M_{\odot}$. This can be attributed
to the fact that in case (b) the clusters in the former mass range have 
lower S-Z fluxes due to the lower temperatures as compared to clusters 
in case (a). Clusters belonging to the latter mass range are still massive
enough (and consequently, hot enough) to compensate for the lack of 
temperature evolution. Regardless of this effect, it should be emphasized 
that the counts in the high mass range are negligible in comparison with 
the total cumulative counts at $z=0$. A slightly different behaviour is 
apparent in case (a) of the $\chi^{2}_1$ model; here total counts are 
still dominated by clusters in the mass range 
$10^{14}h^{-1}M_{\odot}<M<10^{15}h^{-1}M_{\odot}$, although differences 
are not as pronounced as in the Gaussian case. While counts in the 
lower-mass range in the Gaussian case are higher by a factor of 
$\sim 30-40$ than the corresponding counts in the higher-mass range, here 
a mere factor of $\sim 3$ is seen. In case (b) the results are even 
more interesting, since the total counts are dominated by high-mass 
clusters, lying at $0\lesssim z\lesssim 3.5$. The cumulative counts at 
the lowest redshifts consist of a slightly higher contribution from the 
high-mass range and are larger by a factor $\sim 1.3$ than the 
corresponding contribution from mass range 
$10^{14}h^{-1}M_{\odot}<M<10^{15}h^{-1}M_{\odot}$. Differences in total 
counts are large; in the Gaussian model these are $\sim 4000$ and 
$\sim 3000$ in cases (a) and (b), respectively, as compared to levels 
of $\sim 30,000$ and $\sim 10,000$ in the corresponding cases of the 
$\chi^2_1$ model. 

\begin{figure*}
\centering
\epsfig{file=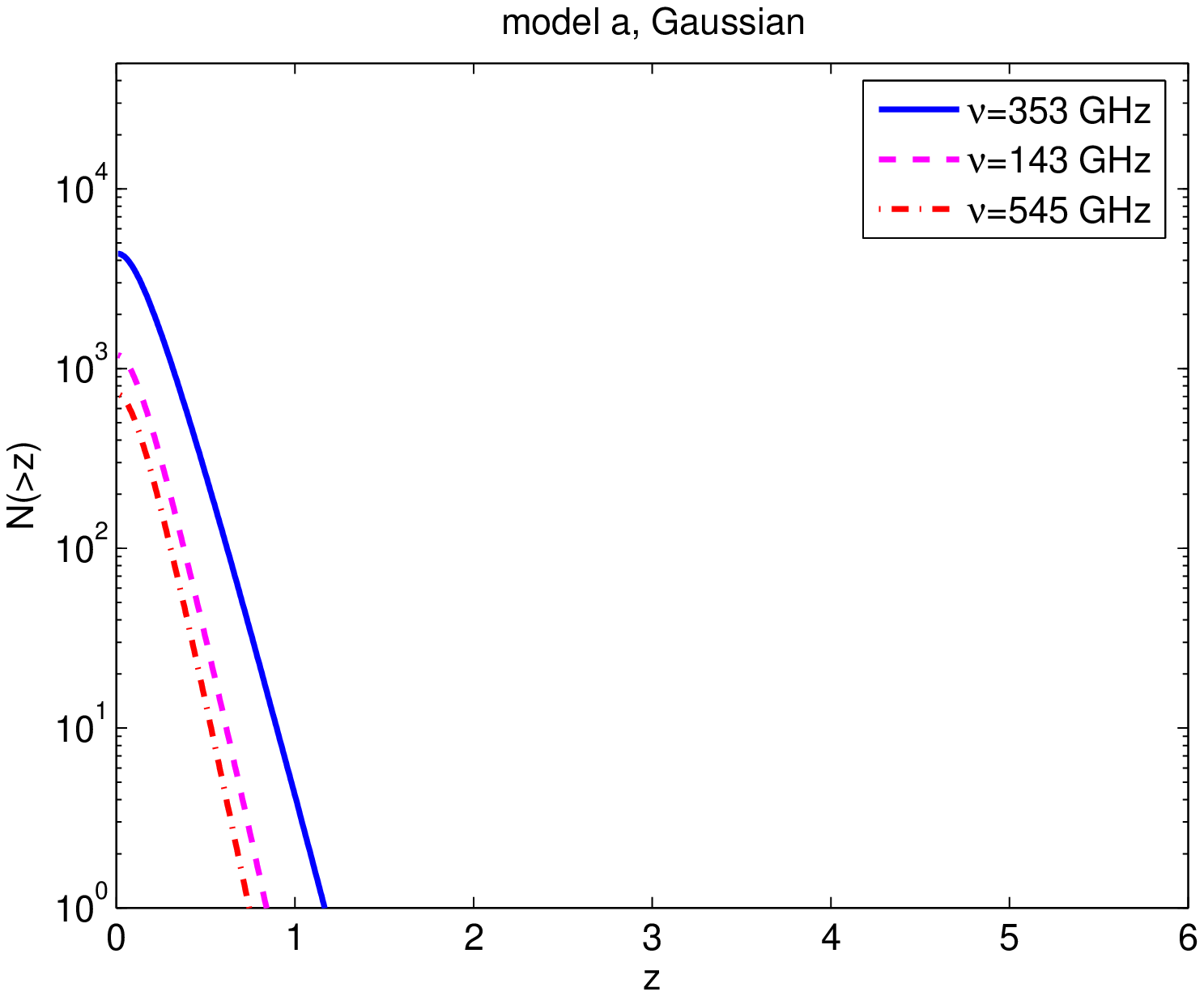, height=7.cm, width=8cm,clip=}
\epsfig{file=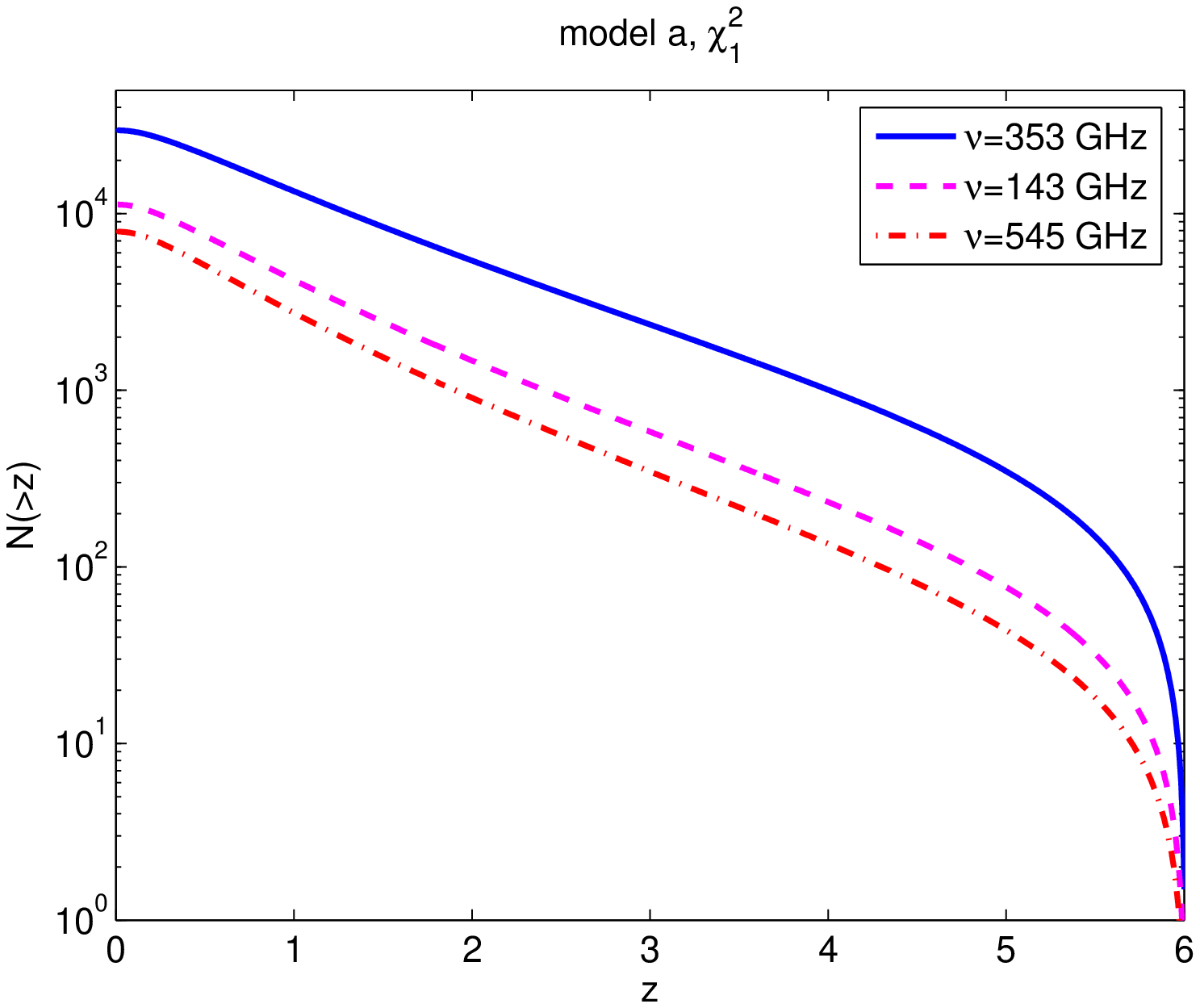, height=7.cm, width=8cm,clip=}
\epsfig{file=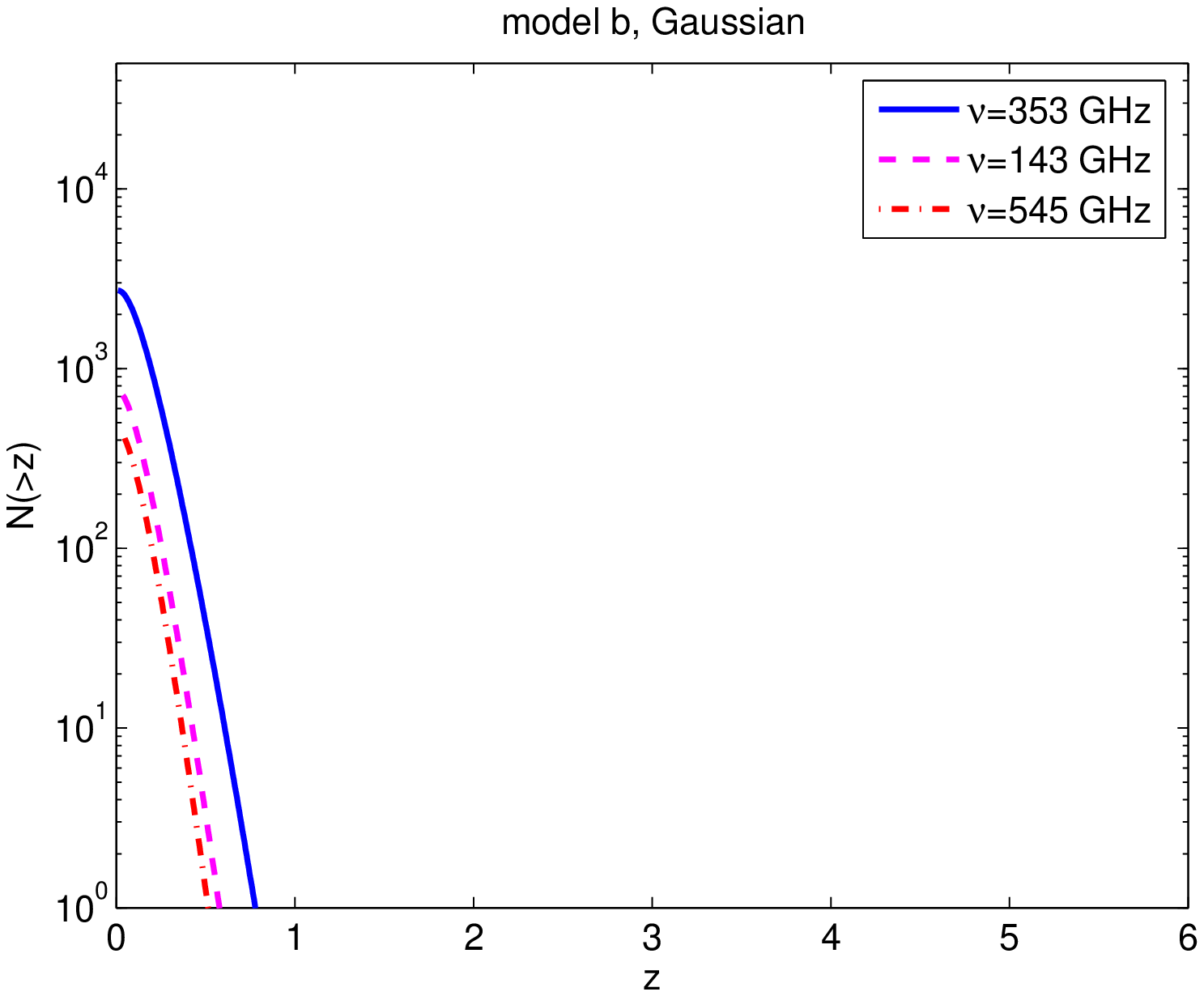, height=7.cm, width=8cm,clip=}
\epsfig{file=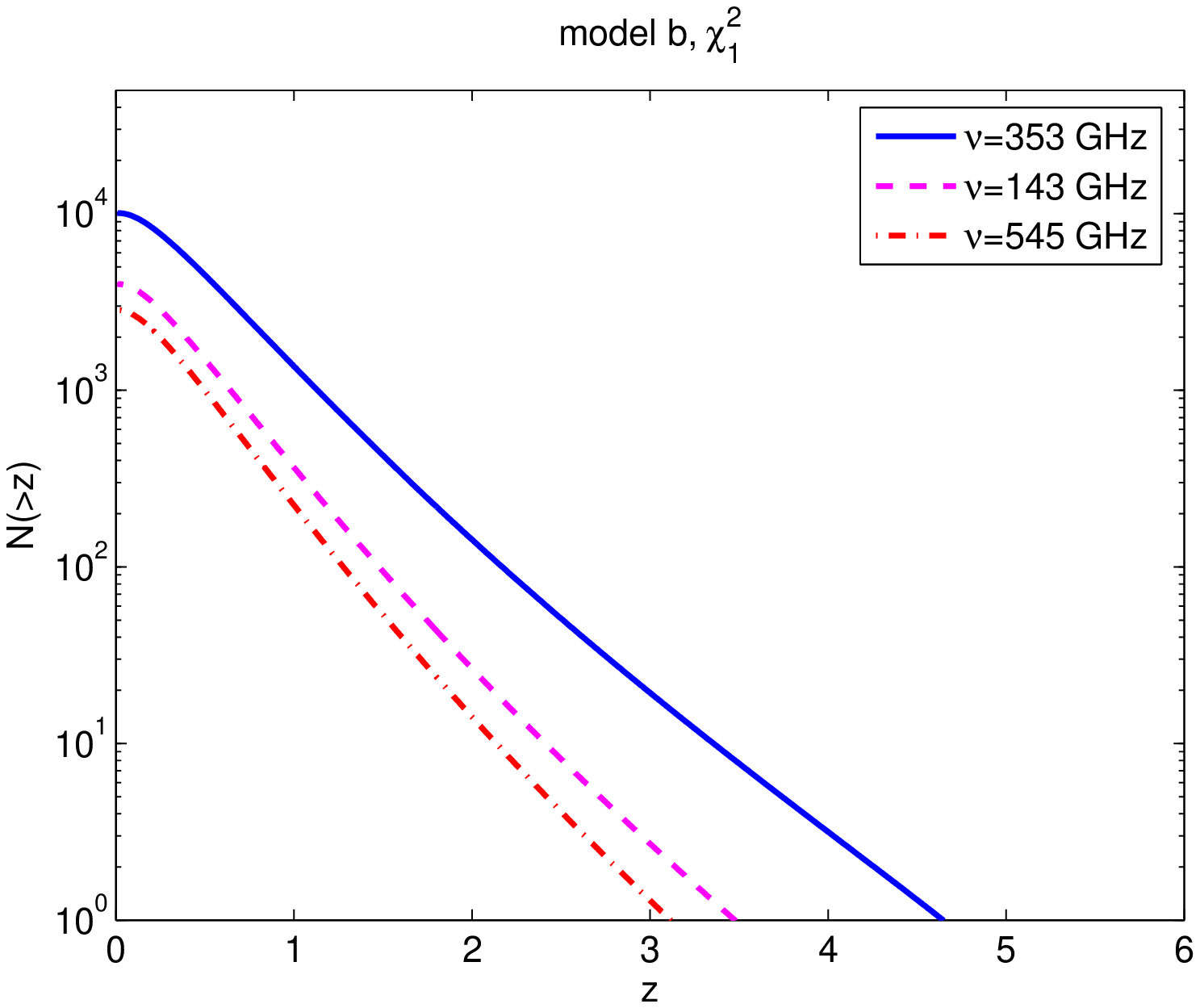, height=7.cm, width=8cm,clip=}
\caption{Cumulative S-Z counts as a function of redshift. Number counts 
are shown for the three specified frequency channels of the HFI 
instrument on the \textit{Planck} satellite.} 
\label{fig:nctz1}
\end{figure*}

Fig.~\ref{fig:nctz1} describes cumulative number counts for two more 
frequency channels, $\nu=143$ GHz and $\nu=545$ GHz, in addition to 
$\nu=353$ GHz which was used above. In the non-relativistic electron 
velocity limit the (thermal component of the) S-Z spectral function - 
which is essentially independent of the gas temperature - is $g(x)$ 
(eq. 19). At these frequencies, $g(x)=-4, 3.2, 6.7$, respectively, 
so our results for $\nu=353$ GHz can be easily scaled accordingly. 
Adopting the same value of the limiting flux ($30 mJy$), we find 
that values of the ratios of number counts at $\nu=353$ GHz and 
$\nu=545$ GHz are larger in the Gaussian than in the $\chi^2_1$ 
model. This can clearly be attributed to the significantly higher 
population of clusters with flux exceeding the limit in the 
$\chi^2_1$ model. Note that an accurate calculation of the S-Z 
intensity change necessitates a relativistic calculation (Rephaeli 
1995) which yields a more complicated expression for the 
{\it temperature-dependent} spectral function (for which there are a few 
analytic approximations; see, e.g., Shimon \& Rephaeli 2004, Itoh 
\& Nozawa 2004). The deviation from the \nrel calculation can amount 
to few tens of percent for typical temperatures in the range 5-10 
keV. Since our main focus here is a comparison between predictions 
of the two density fields for quantities that are integrated over 
the cluster population (rather than an accurate description of the 
effect in a given cluster), it suffices to use the much simpler 
function $g(x)$. 

We may summarize the results of this subsection as follows: 
total count levels are significantly higher in the $\chi^2_1$ model, 
roughly by factors of $\sim 7.5$ and $\sim 3$ in cases (a) and (b), 
respectively. The Contribution of high-mass clusters to the total 
cumulative counts are practically negligible in the Gaussian case, 
but important in the non-Gaussian model, particularly so in case (b) 
where the lower-mass clusters fail to generate sufficient S-Z flux 
by virtue of their lower temperatures with respect to the evolving 
temperature scenario in case (a). In addition to this, the counts 
are more broadly distributed in redshift space in the non-Gaussian 
model due to the higher abundance of clusters at high redshifts.

As in the section describing the S-Z power spectrum, a comparison
with published works on S-Z number counts is pertinent. Springel et al. 
(2001) find $\sim 0.6$ clusters per square degree for an observation
frequency of $150\,GHz$ and a flux limit similar to ours, $30\,mJy$. This
translates into all-sky counts of $\sim 2.5\sim 10^4$ clusters, whereas
we find $\sim 4\cdot 10^{3}$. Holder et al. (2000) report a prediction
of $\sim 25$ clusters per square degree, or a total of over $10^6$ 
clusters. However, it is not clear what was the flux limit they assumed. 
In addition, the slightly higher normalization they 
adopted, $\sigma_8=1$, may be partly responsible for the inferred 
high counts. Benson, Reichardt \& Kamionkowski (2002) find $\sim 0.2$
clusters per square degree for \textit{PLANCK} specifics, with observation
frequency $143\,GHz$ and flux limit $30\,mJy$. An all-sky survey is then
expected to yield $\sim 8000$ clusters. While our predictions are lower, 
it should be emphasized that similar to results for the S-Z power spectrum, 
S-Z cluster counts are equally susceptible to uncertainties in 
IC gas modeling and to the specifics of the experimental setting, such 
that a large scatter in the results is unavoidable. Moreover, our main 
concern in this study is the difference between the magnitudes of S-Z 
observables as predicted by two different mass functions, and those 
would certainly persist were we to modify the modeling of the IC gas and 
any other component relevant to the calculation of the effect.  

\subsection{A comparison between the $\chi^2_1$ and $\chi^2_2$ cases}

It is useful to examine the sensitivity of the results corresponding
to the $\chi^2_m$ model to the number of fields added in quadratures, $m$. 
We have repeated the numerical calculations of the S-Z power spectrum
and number counts (and in addition, the 2-point angular correlation function
of S-Z clusters, which is dealt with separately in the next section) with
$m=2$ for case (a). Since the degree of skewness of the initial probability 
density function is reduced with increasing $m$, this will be manifested in 
the mass function as well, and necessitates a different normalization 
parameter $\sigma_8$. For $m=2$ it is found that in order to reproduce the 
observed cluster abundance at $z=0$, $\sigma_8=0.77$, instead of the lower 
value of $\sigma_8=0.73$ in the $m=1$ model. 

\begin{figure*}
\centering
\epsfig{file=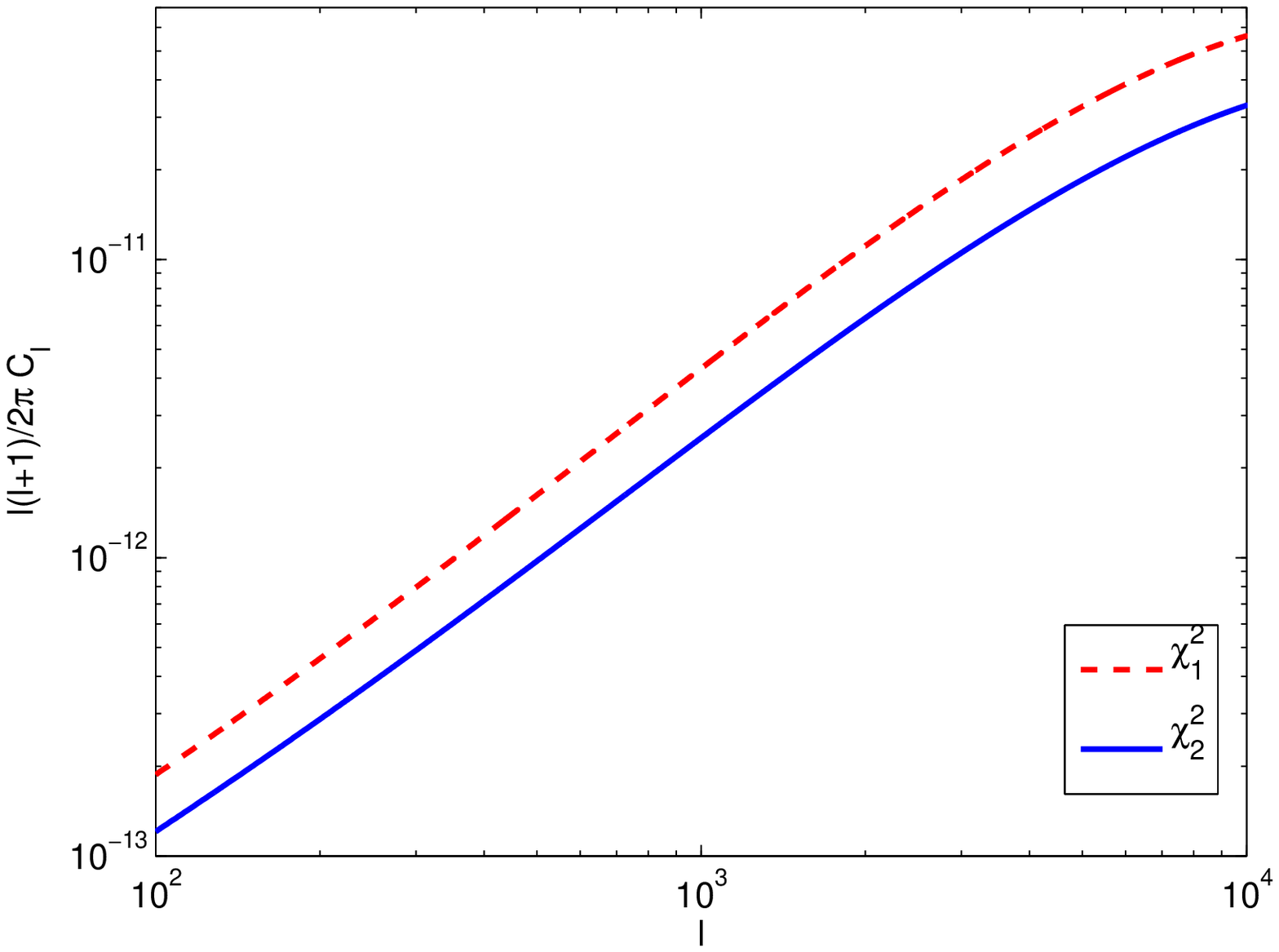, height=7.cm, width=8cm,clip=}
\epsfig{file=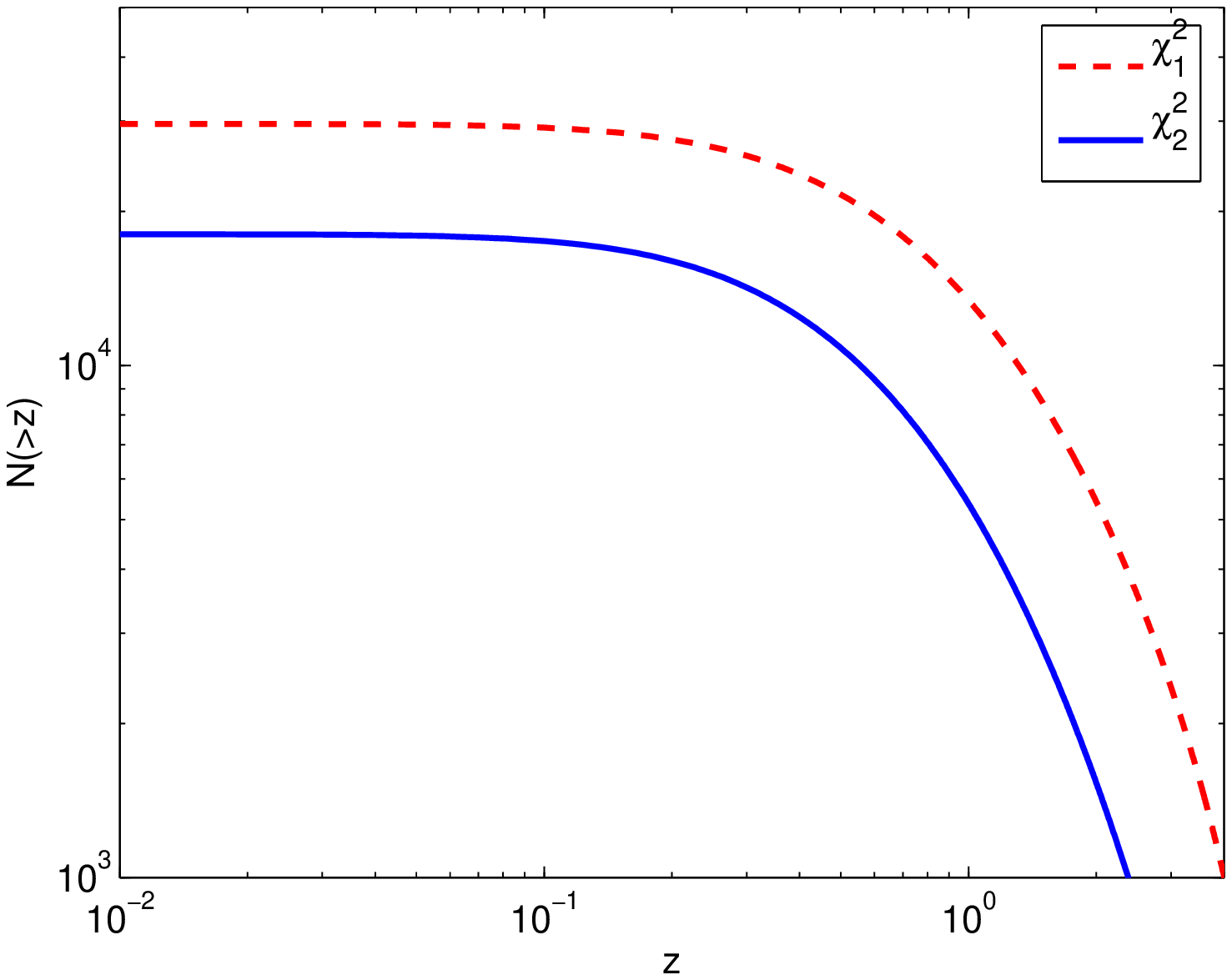, height=7.cm, width=8cm,clip=}
\caption{A comparison between S-Z power levels (left-hand panel) and number 
counts (right-hand panel) for a $\chi^2_m$-distributed density field with
$m=1$ and $m=2$ is shown. 
Calculations were carried out for case (a).}
\label{fig:m12}
\end{figure*}

Results of the calculations are illustrated in Fig.~\ref{fig:m12}. The 
differences are quite significant in both plots; power levels in the $m=1$ 
case are higher by a factor of $\sim 1.5-1.8$ across the multipole range of
$100\le\ell\le 10000$. It is the highest-mass clusters that are most 
susceptible to the high-mass tail of the mass function, which is strongly 
dependent on the parameter $m$. In particular, this is most noticeable at 
high redshifts where the predictions of the mass function for the massive 
clusters in the $m=1$ and $m=2$ cases vary greatly. Since these distant 
clusters have small angular sizes, it is not surprising that the ratio 
between the two curves increases with increasing multipole, reflecting 
the relative higher abundance of clusters with small apparent sizes for 
$m=1$. This is further established by noting that the number counts 
ratio for $m=1$ and $m=2$ are highest at high redshifts: While the 
ratios are $\sim 6$ and $\sim 3.5$ at $z \sim 4$ and $z \sim 2$, 
respectively, the cumulative counts for $m=1$ at $z=0.1$ are 
approximately $\sim 1.5$ times as high as those obtained for $m=2$. 
Clearly, very massive clusters at high redshifts are more common
in the model with $m=1$ than in that with $m=2$.

\section{Angular Correlation Function of S-Z clusters}

The angular 2-point correlation function (hereafter A2PCF) is expected to 
be steeply dependent on the mass function, even more strongly than the 
angular power spectrum and cluster number counts, due to the 
\emph{quadratic} dependence on the cluster abundance. The A2PCF of S-Z 
clusters has been studied by Diaferio et al. (2003), who explored the 
dependence of the expected number of cluster pairs on the biasing 
relation between the cluster and the mass distribution, and Mei \& 
Bartlett (2003), who studied the feasibility of alleviating the 
degeneracy between $\Omega_m$ and $\sigma_8$ by means of combining S-Z 
cluster counts with their angular correlation function. The requisite 
data will become available when targeted S-Z cluster surveys are carried 
out. As pointed out by Mei \& Bartlett, the scientific yield from the 
A2PCF lies in the fact that S-Z cluster catalogs, on their own, may 
constitute an important probe of cosmological models and cluster 
properties, even without the detailed follow-up observations of 
individual clusters. Here we calculate the A2PCF of S-Z clusters for 
the Gaussian and $\chi^2_1$ mass functions and assess its diagnostic 
use to distinguish between the two models.

\subsection{Formalism}

The angular correlation function can be calculated by equating the 
surface density of S-Z clusters with their integrated spatial density:
\begin{eqnarray}
& &d\Omega_1\,d\Omega_2\left[\int_{z}\int_{M} r^{2}(z)\D\frac{dr(z)}{dz}
n\left(M,z\right)\,dz\,dM\right]^2
\left[1+w(\theta)\right]= d\Omega_1\,d\Omega_2\nonumber \\
& &\times\int_{z_1}\int_{z_2}\int_{M_1}\int_{M_2} 
r^{2}(z_1)\D\frac{dr(z_1)}{dz_1}r^{2}(z_2)\D\frac{dr(z_2)}{dz_2}
\,n_1\,n_2\left[1+\xi\left(M_1,M_2,z_1,z_2,R\right)
\right]dz_1\,dz_2\,dM_1\,dM_2,
\label{eq:szcormas}
\end{eqnarray}
where $\xi\left(M_1,M_2,z_1,z_2,R\right)$ is the spatial 2-point 
correlation function of clusters with masses $M_1$ and $M_2$, located at 
redshifts $z_1$ and $z_2$, and separated by physical distance $R$, and 
$n_i\equiv n(M_i,z_i)$. As in the calculation of the number counts, the 
lower limits of the integrals are set so as to yield S-Z fluxes that 
exceed the detection limit. Since the uncorrelated surface density term 
on the left-hand side of equation~(\ref{eq:szcormas}) is identical with 
the integrated spatial density of the uncorrelated term on the 
right-hand side, the angular correlation function can be evaluated as
\begin{eqnarray}
w(\theta)=\D\frac{\D\int_{z_1}\int_{z_2}\int_{M_1}\int_{M_2} 
r^{2}(z_1)\D\frac{dr(z_1)}{dz_1}r^{2}(z_2)\D\frac{dr(z_2)}{dz_2}
\,n_1\,n_2\,\xi\left(M_1,M_2,z_1,z_2,R\right)
\,dz_1\,dz_2\,dM_1\,dM_2}{\left[\D\int_z\int_M r^{2}(z)
\D\frac{dr(z)}{dz}\,n(M,z)\,dz\,dM\right]^2}.
\label{eq:angcf1}
\end{eqnarray}
It remains to specify the spatial correlation function of clusters, $\xi$. 
This can be done either by employing the observational correlation function 
of clusters (e.g. Bahcall \& Soneira 1983), or by regarding it as the biased
counterpart of the correlation function of dark matter, $\xi_{dm}$. We 
choose to employ the second method, in which the cluster correlation 
function is commonly written as 
\begin{equation}
\xi(M_1,M_2,z_1,z_2,R)=b(M_1,z_1)b(M_2,z_2)\xi_{dm}(R,z),
\label{eq:biramp}
\end{equation}
where $b$ is the linear bias factor.
Making use of the global isotropy of the universe, we can express the 
correlation function as the Fourier-transformed power spectrum of 
dark matter:
\begin{equation}
\xi_{dm}(R,z)=\int\int\int e^{i\overline{k}\cdot\overline{R}}
P(\overline{k})\,d^{3}\overline{k}=\frac{A}{2\pi^2}\int_0^{\infty}
k^{n+2}T^{2}(k)\frac{\sin{kR}}{kR}\,dk.
\label{eq:xim}
\end{equation}
The simplest form for the bias factor is (Matarrese et al. 1997, Catelan et al. 
1998) 
\begin{equation}
b(M,z)=1+\frac{\delta_c}{\sigma^{2}_M(z)}-\frac{1}{\delta_c}.
\label{eq:bmz}
\end{equation}
Note that more accurate formulae for the bias factor exist, which take into 
account non-spherical collapse (Sheth, Mo \& Tormen 2001) and nonlinear clustering 
effects (Peacock \& Dodds 1996); we will limit ourselves here to the 
simplest case, since our main objective is this study is a comparative 
evaluation of the predictions of two mass functions (rather than their  
respective accuracy).

To proceed, it is necessary to write down the physical distance $R$ between 
the members of a cluster pair, separated by an angle $\theta$. In doing so 
it is very convenient to take advantage of the fact that the observational
spatial correlation function falls off rapidly with distance, 
$\xi(R)\sim R^{-1.8}$ (e.g. Bahcall \& Soneira 1983); a corresponding trend 
can be inferred from equation~(\ref{eq:xim}), although in this case it is 
obviously different from a power law. Consequently, correlated clusters 
must lie at low angular separations and at approximately the same redshift. 
The separation can be written as
\begin{equation}
R=\sqrt{[r(z_1)-r(z_2)]^2+(d_A\theta)^2},
\end{equation}
where $r$ and $d_A$ denote the radial and angular diameter distance, 
respectively. Substituting $z\equiv(z_1+z_2)/2$ and $u\equiv z_1-z_2<<1$
yields
\begin{equation}
r(z_1)-r(z_2)=r(z_2+u)-r(z_2)\approx r(z_2)+\frac{dr}{dz_2}(z_2)\cdot 
u-r(z_2)\approx u\cdot\frac{dr}{dz}, 
\end{equation}
and the separation simply becomes
\begin{equation}
R=\sqrt{\left[u\cdot\frac{dr}{dz}\right]^2+\left[d_A(z)\cdot\theta\right]^2}.
\label{eq:sep}
\end{equation}
Substituting this in the numerator of equation~(\ref{eq:angcf1}), together 
with the spatial correlation function of equation~(\ref{eq:biramp}), we 
have 
\begin{equation}
r^{2}(z_1)\D\frac{dr(z_1)}{dz_1}r^{2}(z_2)\D\frac{dr(z_2)}{dz_2}\approx
r^4(z)\left(\frac{dr}{dz}\right)^2 du\,dz,
\end{equation}
so that we can now write 
\begin{eqnarray}
& &
w(\theta)=\D\int_{z}\int_{M_1}\int_{M_2} 
r^{4}(z)\left[\D\frac{dr(z)}{dz}\right]^2
\,n(M_1,z)\,n(M_2,z)\,b_1(M_1,z)\,b_2(M_2,z)
\,dM_1\,dM_2\,dz \nonumber \\
\nonumber \\
& &\times\int_{-\infty}^{\infty}\xi\left[\sqrt{[\D\frac{dr}{dz}\cdot u]^2+
[d_A(z)\cdot\theta]^2},z\right]du\times
\left[\D\int_z\int_M r^2(z)\D\frac{dr(z)}{dz}\,n(M,z)\,dz\,dM\right]^{-2}.
\label{eq:bigint1}
\end{eqnarray}
The integration over the spatial correlation function may be performed by
substituting the cluster separation $R$ (equation~\ref{eq:sep}) in the 
dark matter correlation equation~(\ref{eq:xim}), which is in turn 
substituted in the integral over $u$ in equation~(\ref{eq:bigint1}):
\begin{equation}
\frac{A}{2\pi^{2}}\int_0^{\infty}k^{n+2}T^2(k)\,dk
\int_{-\infty}^{\infty}\frac{\sin{k
\sqrt{\left(\frac{dr}{dz}\cdot u\right)^2+(d_A(z)\cdot\theta)^{2}}}}
{k\sqrt{\left(\frac{dr}{dz}\cdot u\right)^2
+(d_A(z)\cdot\theta)^2}}\,du,
\label{eq:acor}
\end{equation}
where the last expression was obtained by changing the order of 
integration. The integral over $u$ can be performed using the following 
change of variables: 
\begin{equation}
x\equiv\frac{{\sqrt{\left(\frac{dr}{dz}\cdot u\right)^2+
(d_A(z)\cdot\theta)^2}}}{d_A(z)\cdot\theta},
\end{equation}
which results in an integral that can be put in a closed form 
\begin{equation}
\frac{2}{k(dr/dz)}\int_{1}^{\infty}\frac{\sin{[k\,x\,d_A(z)\theta]}\,dx}
{\sqrt{x^2-1}}=\frac{\pi}{k(dr/dz)}J_0[kd_A(z)\theta],
\end{equation}
where $x$ was defined above and $J_0$ is a Bessel function of the first 
kind and order zero. Expression~(\ref{eq:acor}) then finally becomes
\begin{equation}
\frac{A}{2\pi(dr/dz)}\int_0^{\infty}k^{n+1}T^2(k)J_0[kd_A(z)\theta]\,dk.
\label{eq:dmappw}
\end{equation}
This integral can be easily evaluated numerically.

\subsection{Results}

\begin{figure*}
\centering
\epsfig{file=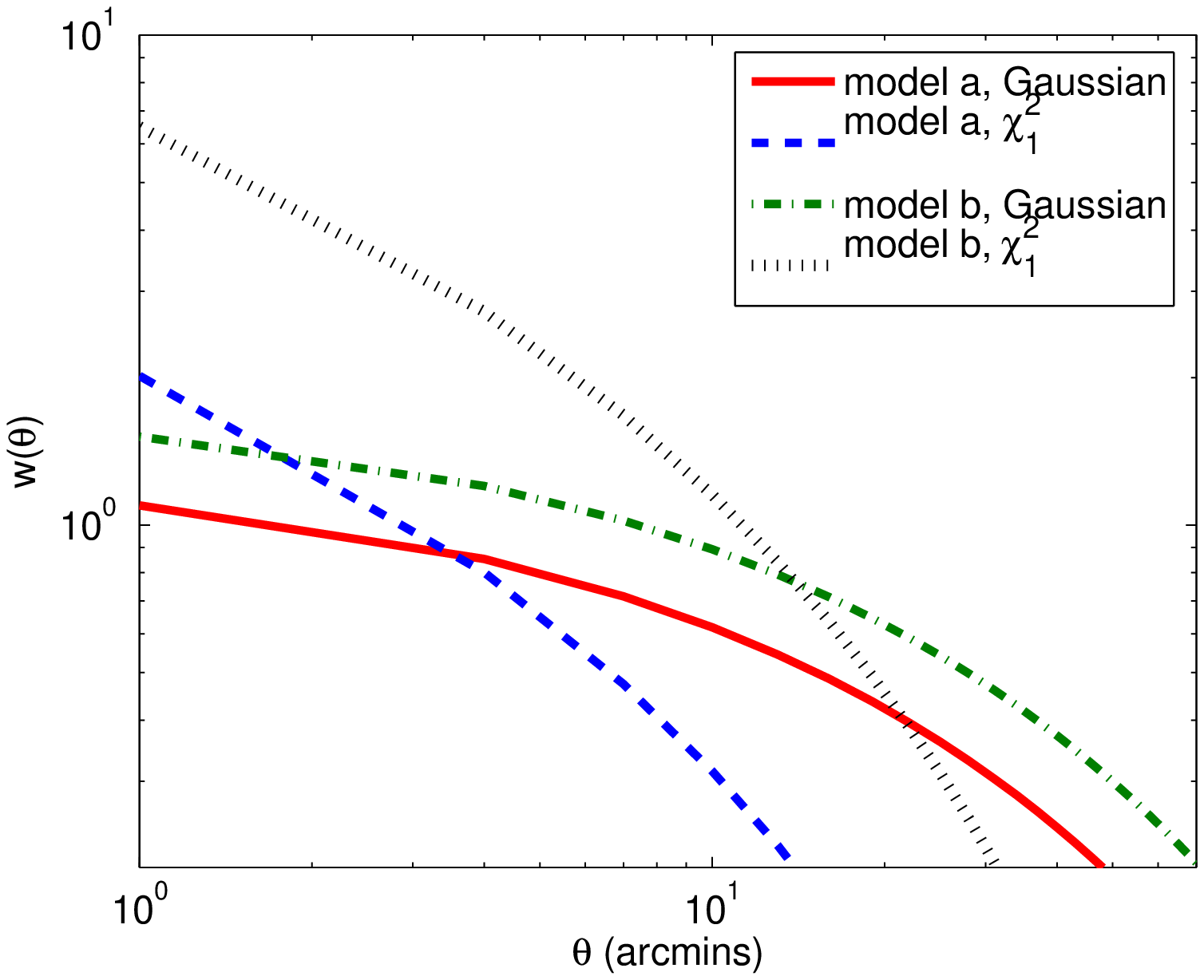, height=7.cm, width=8cm,clip=}
\epsfig{file=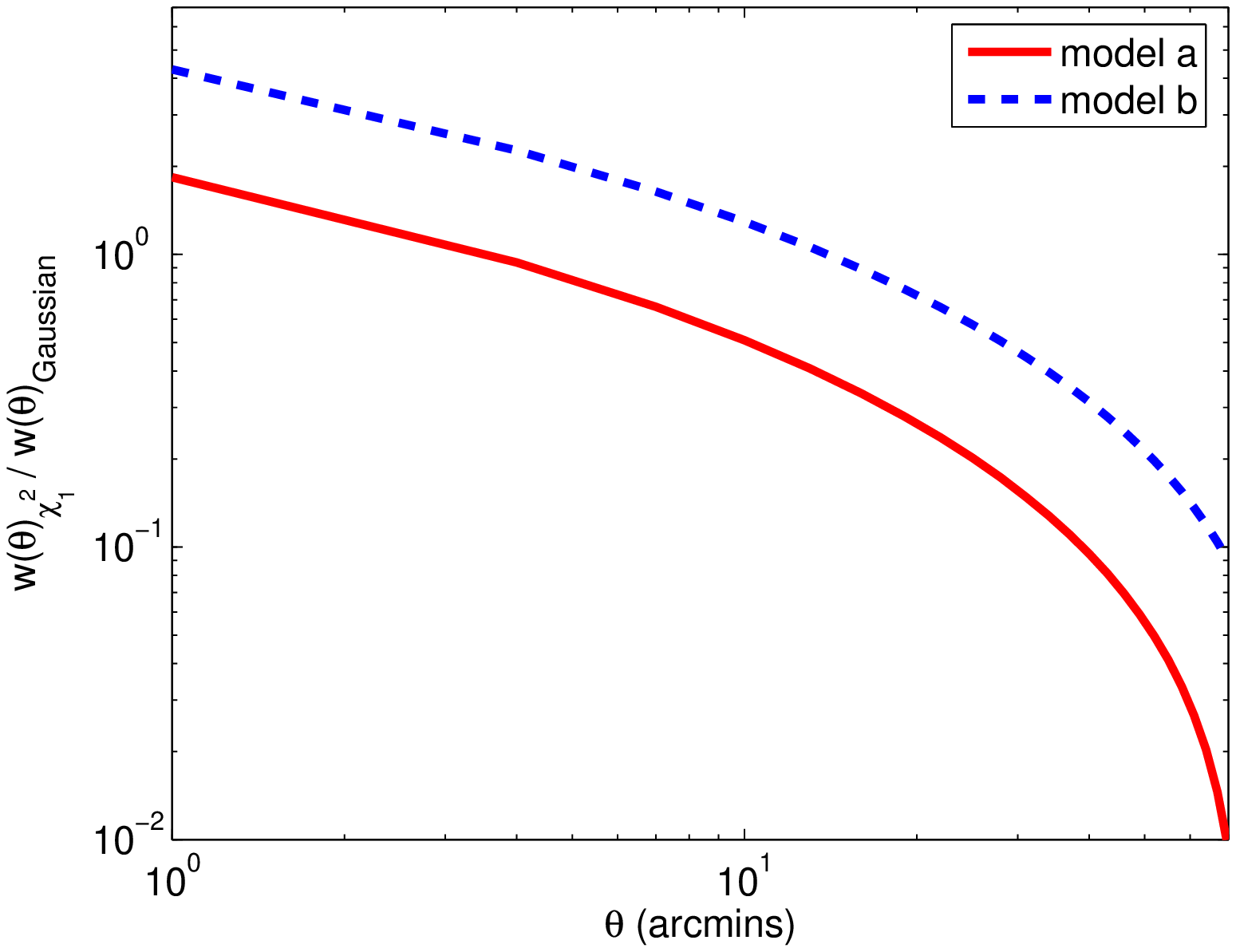, height=7.cm, width=8cm,clip=}
\caption{The A2PCF calculated for the Gaussian and $\chi^2_1$ models for 
cases (a) and (b). The left-hand panel describes the correlation levels, 
whereas the right-hand panel depicts the ratios between the $\chi^2_1$ 
and Gaussian models.}
\label{fig:szc}
\end{figure*}

The A2PCFs in cases (a) and (b) in the Gaussian and $\chi^2_1$ models 
are plotted in Fig.~\ref{fig:szc}. Their shapes are dictated by the 
functional form of the spatial correlation function, which is dominated by 
the Bessel function of the first kind and order zero appearing in 
equation~(\ref{eq:dmappw}). Referring first to the differences between
the predictions of cases (a) and (b), it can be seen that in both the Gaussian
and $\chi^2_1$ models the correlation levels are higher in case (b). The 
reason for this is easily explainable noting that case (b) is biased towards
higher-mass clusters owing to its lack of temperature evolution; as clusters
do not become hotter with increasing redshift as in case (a), only very 
massive clusters generate detectable S-Z signals. These higher-mass clusters 
are more strongly correlated, as indicated by the the bias term, 
equation~(\ref{eq:bmz}), by virtue of its specific dependence on the mass 
variance $\sigma(M)$, a monotonically decreasing function of mass $M$. 
Thus, the bias factor is a monotonically increasing function of $M$, giving
rise to stronger correlations among high-mass cluster populations. 

In order to explain the differences between the predictions of the 
$\chi^2_1$ and Gaussian models in either case (a) or (b), it will first be 
useful to give a qualitative explanation of the shape of the A2PCF curve. 
We may think of an effective maximum physical separation for which (positive) 
correlation between two clusters is still possible. At low angular 
separations the scale corresponding to this limiting distance can be 
subtended at relatively high redshifts, so contribution to the A2PCF at 
this angle is possible from low as well as high redshifts. With increasing 
angular separation the corresponding scale can only be subtended at lower 
redshifts, so that the main contribution to the correlation is generated 
locally. Since the $\chi^2_1$ model yields higher cluster populations 
(and particularly so of high-mass clusters) at high redshifts, it is natural 
to expect higher correlation levels than in the corresponding Gaussian case, 
where the cluster population is limited to lower redshifts, and also 
relatively limited in massive clusters. As can be seen in the right-hand 
panel of Fig.~\ref{fig:szc} at angular separations of $\theta\gtrsim 3'$ 
and $\theta\gtrsim 10.5'$ in cases (a) and (b), respectively, the 
correlation levels in the $\chi^2_1$ model fall below those of the 
Gaussian model. Since the redshift range available for contributing to 
the angular correlation function reduces with increasing angular separation, 
this trend is indicative of a diminished contribution of relatively 
low-redshift clusters in the $\chi^2_1$ model. In fact, this is indeed the 
case, and can be easily realized by inspection of the upper and lower-right 
hand panels of Figs.~\ref{fig:clz} and~(\ref{fig:nctz}), where it has 
been shown that a substantial contribution to the power spectrum levels at 
high multipoles and to the cumulative number counts originates in clusters 
lying at high redshifts ($z>1$). As explained above, at such high redshifts 
clusters are not strongly correlated at high angular separations by virtue 
of their mutual large physical distances. On the other hand, in the Gaussian 
model the corresponding contribution comes almost exclusively from 
low-redshift clusters, which may manifest relatively strong correlations 
even at higher angular separations. Clearly, this effect is also 
responsible for the steeper shape of the A2PCF curve in the non-Gaussian 
models. 

\section{Discussion and Conclusion}

The main objective of this study has been to compare predictions of S-Z 
observables for two mass functions which differ mainly at the high mass 
end. A $\chi^2_1$ distributed density fluctuation field is characterized 
by a longer tail of high density fluctuations with respect to the Gaussian 
case, and is capable of generating more high-mass clusters at higher 
redshifts, owing to the denser population of high density fluctuations 
(or peaks) which can collapse at earlier times. S-Z power spectrum, 
number counts, and 2-point angular correlation function all prove to 
yield significantly different results if calculated using a mass function 
derived from a Gaussian or $\chi^2_1$ primordial density field. We have 
demonstrated that levels of the power spectrum in the latter model 
are higher by about an order of magnitude than in the former model, 
depending on the IC temperature scaling with the cluster mass. This 
was explicitly shown to take effect in light of the larger contribution 
of high-mass clusters and high redshift clusters to the total power 
in the non-Gaussian model. The larger contribution from distant 
clusters is also responsible for the higher multipoles at peak power. 

We have shown that S-Z power spectrum levels are contributed by different 
cluster populations in the Gaussian and $\chi^2_1$ models; in the 
former model, most of the power originates in clusters lying at $0<z<1$, 
with the exception of the higher multipoles for which a non-negligible 
amount of power comes from $1<z<2$. In contrast, in the latter model 
more power is contributed from higher redshift clusters, beginning at 
multipole $\sim 3000$. On the other hand, most of the power is 
contributed by clusters lying in the mass range $10^{14}<M<10^{15}$ 
$M_{\odot}$. This holds for both models in the interesting multipole 
ranges. 

The power spectrum is quite sensitive also to cluster internal properties, 
as is demonstrated in part through the modification of the cluster 
temperature scaling with mass, and by the dependence on the cluster 
core radius. These reflect the uncertainty of the actual scalings 
due to observational scatter, lack of high-redshift cluster data, and 
insufficient theoretical information. Plausible scatter of $10-20\%$ in 
these scalings results in significant differences in levels of power, 
albeit to a considerably lower degree than those produced by replacing 
the Gaussian mass function with the $\chi^2_1$ mass function. Number counts 
are also significantly higher in the $\chi^2_1$ model (up to an order of 
magnitude), reflecting the higher population of high-redshift, massive 
clusters, capable of generating sufficient S-Z flux to be detected by 
upcoming experiments. These results obviously depend on the flux 
detection limit of the experiment, and are likely to become more 
pronounced with increasing limiting flux, owing to the high-mass 
cluster bias. 

The 2-point angular correlation function of S-Z clusters is potentially 
an additional observable for which the Gaussian and $\chi^2_1$ models may 
result in substantially different predictions. Depending on the 
temperature scaling with redshift, it has been shown that the 
correlation level at the lowest angular separations ($\sim 1'$) is 
higher by a factor $\sim 2$ and $\sim 4$ in the $\chi^2_1$ model. 
On the other hand, at larger angular separations correlation levels 
are higher in the Gaussian model. This characteristic behaviour leads 
to a steeper functional form in the $\chi^2_1$ model. Although not 
explicitly demonstrated in this study, the flux limit of the experiment 
is, of course, a major factor in determining the correlation levels, 
due to the fact that higher flux limits are biased towards more massive 
clusters, whose bias parameters are larger. 

An additional relevant property which may be manifested differently 
in Gaussian and non-Gaussian density fields is cluster (or halo) 
formation time. Clusters are likely to form earlier in (positively skewed)
non-Gaussian fields owing to the higher probability of finding an 
overdensity exceeding the critical density for collapse with respect
to the corresponding probability in a Gaussian density field. This may 
bear consequences for, e.g., the concentration parameter of haloes, a
key parameter of the NFW dark matter profile (Navarro, Frenk \& White 1995).
The concentration parameter is expected to assume higher values with
earlier formation time by virtue of the higher background density. 
In this context the recent result from a lensing analysis of the cluster 
A1689 (Broadhurst et al. 2005) is of particular interest. A very steep mass 
profile was deduced in this analysis, one that is fit by an extremely 
high concentration parameter of $c=13.7^{+1.4}_{-1.1}$. This value seems 
to be at a considerable variance with theoretical predictions. Observations 
of two other clusters yield $c\sim 12$ for MS 2137-23 (Gavazzi et al. 2003), 
and $c\simeq 22$ and $c\simeq 4$ for the main and secondary clumps in the 
cluster C1 0024 (Kneib et al. 2003). Although perhaps not very meaningful 
from a statistical point of view, these results may be indicative of 
earlier formation times, which are favoured in positively-skewed 
non-Gaussian fields. 

The temperature of the IC gas of virialized clusters may also be affected
by the cluster formation time. In conventional
temperature-mass relations it is the observed redshift of a cluster
that determines its temperature. In this \emph{recent formation} approach,
the relevant redshift is the observation redshift. Clearly, this picture
does not address the merging process that gives rise to the clusters we
observe today. In the merger picture the formation redshift, $z_{f}$,
may differ significantly from the cluster observed redshift, $z_{obs}$,
and the predicted temperatures will vary accordingly. 

The temperature and concentration parameter dependence on formation time
are likely to influence S-Z observables; higher IC gas temperatures (stemming
from earlier formation times) may give rise to an intensification of the effect, 
whereas, if one assumes that IC gas approximately traces the dark matter
distribution, higher concentration parameters may bring about transfer of 
power to higher multipoles, in light of the resulting steeper gas density 
profile. 

We have conducted a preliminary study of the probability distribution function
(PDF) of formation times within the framework of Gaussian and $\chi^2_2$ density
fields. This initial work is based on the excursion set formalism and its derivatives
(Bond et al. 1991, Lacey \& Cole 1993). We note that the definition of the formation 
time used here is somewhat arbitrary; clusters were assumed to have formed once they 
assembled a fraction 
of either $f=0.5$ or $f=0.75$ of their current masses. We have found that on 
most relevant mass scales clusters form at higher redshifts in the 
$\chi^2_2$ model. Defining ``effective'' cluster temperatures and concentration 
parameters by averaging these two quantities at specified formation 
redshifts over the PDF of formation times, we learn that temperatures and 
concentration parameters are on average higher in the $\chi^2_2$ model for 
masses $10^{13}-10^{15}$ $M_{\odot}$. 
At higher masses the differences are practically negligible. Also, with 
increasing observation redshift, the differences become gradually less 
pronounced. This is related to the fact that the time range available for 
formation is reduced with increasing observation time, since the cluster must 
have formed at a redshift exceeding the observation redshift. 

\begin{figure*}
\centering
\epsfig{file=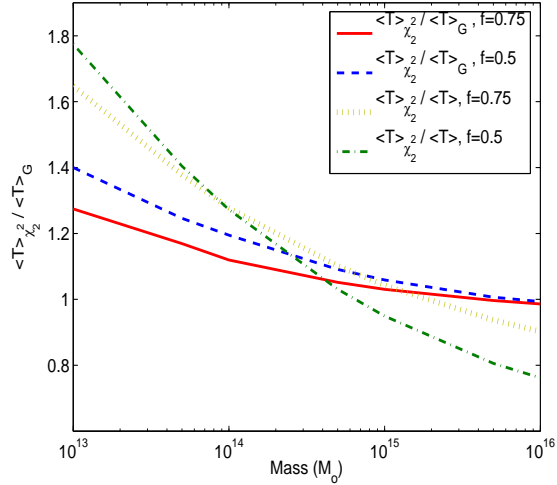, height=7.cm, width=8.cm, clip=}
\caption{Ratios of effective temperatures are shown. The
average temperature predicted in the $\chi^2_2$ model is divided by
the corresponding quantity in the Gaussian model for $f=0.5$ and $f=0.75$. 
Also shown is the ratio between the effective temperature in the 
non-Gaussian model and the temperature predicted by the temperature-mass 
relation provided in equation (13).}
\label{fig:avtem}
\end{figure*}

As an example of our preliminary results we include Fig.~(\ref{fig:avtem}), 
which compares the effective IC gas temperature (as defined above) in 
the $\chi^2_2$ and Gaussian models as a function of cluster mass, for $f=0.75$ 
and $f=0.5$. Also shown are the respective temperature ratios between the 
$\chi_2^2$ predictions and those given by equation~(\ref{eq:tempsc}).
Clearly, the indicated average temperatures are closely related to the
redshift range at which the formation time PDF attains the highest levels.
For example, the PDFs of the lowest mass scale ($10^{13}M_{\odot}$) peak at
relatively high redshifts, and therefore the weight of the contribution
from the $(1+z)$ factor is rather large. This results in higher average
temperatures. The effect is most pronounced when comparing the
$\chi^2_2$ model with the non-averaged temperature of
equation~(\ref{eq:tempsc}), since for the latter case all clusters are
assumed to have formed at $z_{obs}=0$. Obviously, $f=0.5$ implies earlier
formation times and higher average temperatures for a given (low)
cluster mass. This holds also for the $\chi^2_2$ model, which involves
earlier formation times for this mass range. 
The average temperature predictions in the $\chi^2_2$ and Gaussian
models show that clusters of $10^{13} \, M_{\odot}$ are hotter by factors
$\sim 1.4$ and $\sim 1.27$ for $f=0.5$ and $f=0.75$, respectively. For
$M=10^{14} \, M_{\odot}$ the corresponding factors are $\sim 1.23$ and
$\sim 1.12$, while for $M=10^{15} \, M_{\odot}$ in both cases the factors
are $\sim 1$.

We may therefore conclude as follows: 
a mass function based on a $\chi^2_m$ density fluctuation field gives 
rise to higher populations of massive clusters at relatively high 
redshifts with respect to a mass function derived on the basis of 
Gaussian density fluctuation fields. This generates higher levels of 
S-Z power spectrum and number counts by up to an order of magnitude. 
The 2-point angular correlation function of S-Z clusters is also 
susceptible to the mass function model and predicts higher correlations 
at low angular separations in the non-Gaussian model. At higher
angular separations the correlation levels in a Gaussian field 
dominate. Cluster formation times are generally earlier in the 
non-Gaussian model, and may also give rise to higher temperatures and 
concentration parameters of low-mass clusters observed at relatively 
low redshifts. A more rigorous exposition of the topic of formation
times, their relevance to S-Z observables, and further related 
ramifications will be discussed in a future paper. 

\section{Acknowledgment}

Work at Tel Aviv University is supported by a grant from the Israel
Science Foundation. J.S. gratefully  acknowledges support as a Visiting
Sackler Scholar at the Institute of Advanced Studies, Tel Aviv University.

\section{References}
\def\ref{\par\noindent\hangindent 20pt}
\ref Bahcall N.A., Soneira R.M., 1983, \apj, 272, 627
\ref Bardeen J.M., Bond J.R., Kaiser N., Szalay A.S., 1986,
\apj, 304,15
\ref Benson A.J., Reichardt C., Kamionkowski M., 2002, \mn, 331, 71
\ref Bond J.R., Cole S., Efstathiou G., Kaiser N., 1991, 
\apj, 379, 440
\ref Bond J.R. et al., 2005, \apj, 626, 12
\ref Broadhurst T., Takada M., Umetsu K., Kong X., Arimoto N., 
Chiba M., Futamase T., 2005, \apj, 619, L143-L146
\ref Carlstrom, J.E., Holder, G.P., Reese, E.D. 2002, ARA\& A, 40, 643
\ref Carroll S.M., Press W.H., Turner E.L., 1992, \ar, 30, 499 
\ref Catelan P., Lucchin F., Matarrese S., Porciani C., 1998, \mn,
297, 692
\ref Colafrancesco S., Mazzotta P., Rephaeli Y., Vittorio N., 1997,
\apj, 479, 1
\ref Da Silva A.C., Kay S.T., Liddle A.R., Thomas P.A., Pearce F.R., Barbosa D.,
 2001, \apj 561, L15-L18 
\ref Diaferio A., Nusser A., Yoshida N., Sunyaev R.A., 2003, \mn,
338,433
\ref Gavazzi R., Fort B., Mellier Y., Pell\'{o} R., Daniel-Fort M., 
2003, \aa, 403, 11
\ref Holder G.P., Mohr J.J., Carlstrom J.E., Evrard A.E., Leitch E.M., 2000,
\apj, 544, 629
\ref Itoh N., Nozawa S., 2004, \aa, 417, 827
\ref Kneib J.P. et al., 2003, \apj, 598, 804
\ref Komatsu E. et al., 2003, \apjs, 149, 119
\ref Koyama K., Soda J., Taruya A., 1999, \mn, 310, 1111
\ref Kuo C.L. et al., 2004, \apj, 600, 32   
\ref Kurk J., Venemans B., R\"{o}ttgering H., Miley G., Pentericci L., 
2004, in Plionis M., ed., Multiwavelength Cosmology. Kluwer, Dordrecht, 
preprint (astro-ph/0309675)
\ref Lacey C., Cole S., 1993, \mn, 262, 627
\ref Mason B., Pearson T.J., Readhead A.C.S., Sheperd M.C., Sievers J.L., 
Udomprasert P.S., Cartwright J.K., Farmer A.J., et al., 2003, \apj, 591, 540
\ref Matarrese S., Coles P., Lucchin F., Moscardini L., 1997, \mn,
286, 115
\ref Mathis H., Diego J.M., Silk, J., 2004, \mn, 353, 681
\ref Mei S., Bartlett J.G., 2003, \aa, 410, 767
\ref Miley G.K. et al., 2004, \n, 427, 47
\ref Molnar S.M., Birkinshaw M., 2000, \apj, 537, 542
\ref Mullis C.R. et al., 2003, \apj 594, 154 
\ref Navarro J.F., Frenk C.A., White S.D.M, 1995, \mn, 275, 720
\ref Norman, M.L. 2005, preprint (astro-ph/0511545)
\ref Peacock J.A., Dodds S.J., 1996, \mn, 280, 19
\ref Press W.H., Schechter P., 1974, \apj, 187, 425
\ref Refregier A., Teyssier R., 2002, Phys. Rev. D, 66, 043002
\ref Rephaeli Y., 1995, \ar, 33, 541
\ref Sheth R.K., Mo H.J., Tormen G., 2001, \mn, 323, 1
\ref Shimon M., Rephaeli Y., 2004, \na, 9, 69
\ref Springel V., White M., Hernquist L., 2001, \apj, 549, 681
\ref Vikhlinin A., McNamara B.R., Forman W., Jones C., Quintana J., 
Hornstrup A., 1998, \apj, 498, L21

\bsp

\label{lastpage}

\end{document}